\documentclass{article}
\usepackage[utf8]{inputenc}
\usepackage{graphicx}        % standard LaTeX graphics tool
\usepackage{url}
\usepackage{amsmath}
\usepackage{multirow}
\usepackage{graphicx}
\usepackage{caption}
\usepackage{subcaption}
\usepackage{alltt}
\newtheorem{Definition}{Definition}
\usepackage[algo2e]{algorithm2e}

\usepackage{listings}
\usepackage{color}
\usepackage{colortbl}
\usepackage{appendix}

\definecolor{codegreen}{rgb}{0,0.6,0}
\definecolor{codegray}{rgb}{0.5,0.5,0.5}
\definecolor{codepurple}{rgb}{0.58,0,0.82}
\definecolor{backcolour}{rgb}{0.95,0.95,0.92}

\lstdefinestyle{mystyle}{
    commentstyle=\color{codegreen},
    keywordstyle=\color{magenta},
    numberstyle=\tiny\color{codegray},
    stringstyle=\color{codepurple},
    basicstyle=\footnotesize,
    breakatwhitespace=false,
    breaklines=true,
    captionpos=b,
    keepspaces=true,
    numbers=left,
    numbersep=5pt,
    showspaces=false,
    showstringspaces=false,
    showtabs=false,
    tabsize=2
}

\begin{document}

\title{Evaluating Community Detection Algorithms for Progressively Evolving Graphs}

\author{%%%% First author details
Remy Cazabet$^*$ \\
Univ de Lyon, CNRS, Université Lyon 1, LIRIS,\\ UMR5205 Villeurbanne France%%%%%%% Second author details
\and Souâad Boudebza \\
Université Mohamed Seddik Benyahia de Jijel, \\BP 98 Ouled Aïssa, 18000, Jijel, Algeria
%%%%%%%
\and
%%%%%%% Third author details
Giulio Rossetti \\
Information Science and Technology Institute \\ of the Italian National Research Council}

% \author{{%%%% First author details
% \sc Remy Cazabet},\\[2pt]
% Univ de Lyon, CNRS, Université Lyon 1, LIRIS, UMR5205 Villeurbanne France\\
% {\email{remy.cazabet@gmail.com}}\\[2pt]
% %%%%%%% Second author details
% {\sc Souâad Boudebza }\\[2pt]
% Université Mohamed Seddik Benyahia de Jijel, BP 98 Ouled Aïssa, 18000, Jijel, Algeria\\
% {s_boudebza@esi.dz}\\[6pt]
% %%%%%%%
% {\sc and}\\[6pt]
% %%%%%%% Third author details
% {\sc Giulio Rossetti} \\[2pt]
% Information Science and Technology Institute of the Italian National Research Council\\
% {giulio.rossetti@isti.cnr.it}}

\maketitle

\begin{abstract}
{Many algorithms have been proposed in the last ten years for the discovery of dynamic communities. However, these methods are seldom compared between themselves. In this article, we propose a generator of dynamic graphs with planted evolving community structure, as a benchmark to compare and evaluate such algorithms. Unlike previously proposed benchmarks, it is able to specify any desired evolving community structure through a descriptive language, and then to generate the corresponding progressively evolving network. We empirically evaluate six existing algorithms for dynamic community detection in terms of instantaneous and longitudinal similarity with the planted ground truth, smoothness of dynamic partitions, and scalability. We notably observe different types of weaknesses depending on their approach to ensure smoothness, namely \textit{Glitches}, \textit{Oversimplification} and \textit{Identity loss}. Although no method arises as a clear winner, we observe clear differences between methods, and we identified the fastest, those yielding the most smoothed or the most accurate solutions at each step. }
{Dynamic Networks, Community Detection, Dynamic Communities, Network Generator}
%%%% If classification number provided then
%\\
%2000 Math Subject Classification: 34K30, 35K57, 35Q80,  92D25
\end{abstract}

%\author[1]{Cazabet Rémy}
%\affil[1]{\small{Univ Lyon, UCBL, CNRS, LIRIS UMR 5205, F-69621, Lyon, France \thanks{remy.cazabet@euniv-lyon1.fr}}}

%\begin{document}
%\date{}

%\maketitle

%TODO Aris Anagnostopoulos : Community Detection on Evolving Graphs

\section{Introduction}

Many algorithms have been proposed in recent years to discover evolving communities in dynamic networks. Because few empirical comparisons of them have been conducted, their relative strengths and weaknesses are mostly unknown. A likely reason for the scarcity of such work is the lack of reliable benchmarks to generate synthetic graphs, i.e., an equivalent to the LFR benchmark \cite{lancichinetti2008benchmark} in static settings. Several benchmarks have already been proposed (e.g. \cite{granell2015benchmark,bazzi2016generative,Rossetti2017,sengupta2017benchmark}), but none of them allow to generate a dynamic network corresponding to a scenario of community evolution described by the experimenter.

In this paper, we will focus on the problem of detecting communities in \textit{progressively evolving graphs}, i.e., graphs for which the graph is well defined at any given time, and change at a slow rate. Such graphs are common in human activities \cite{perc2010coevolutionary}, for instance friendships in social networks, physical infrastructures (electricity/transport network, etc.) and physical proximity between individuals captured at a high rate using personal sensors\cite{Genois2018}.

Since reproducibility is paramount in such a work, we provide an open-source implementation of the benchmark generator, algorithm implementations and evaluation scores as a fully documented python library and a notebook allowing to reproduce the results\footnote{Library: \url{https://tnetwork.readthedocs.io/en/latest/}. \\Experiments reproduction: \url{https://tinyurl.com/y7a2lrbz}}.

\subsection{Ship of Theseus and the nature of dynamic communities}

Communities lifecycles --their history, the events they undergo, etc.-- are of utmost importance since they are what makes the difference between static and dynamic community detection. Indeed, two algorithms that agree on what is the best partition for each static graph composing the dynamic network might still disagree on what the corresponding \textit{dynamic} communities are. A good example of this problem is to consider the \textit{ship of Theseus} paradox.

\vspace{0.5cm}

\label{theseus}

The paradox of the ship of Theseus is an ancient thought experiment about the identity of an object evolving through time. It can be formulated as follows:

Let's consider a famous ship, the ship of Theseus, composed of planks, and kept in a harbor as a historical artifact. As time passes, some planks deteriorate and need to be replaced by new ones. After a long enough period, all the original planks of the ship have been replaced. Can we consider the ship in the harbor to still be the same ship of Theseus? If not, at which point exactly did it cease to be the same ship?

Another aspect of the problem arises if we add a second part to the story. Let's consider that the old planks were stored in a warehouse, repaired, and that a new ship, identical to the original one, is built with them. Should this ship, just built out, be considered as the \textit{real} ship of Theseus?

Let's call the original ship A, the ship that stayed in the harbor B, and the reconstructed from original pieces, C.

\begin{figure}
  \centering
    \includegraphics[width=0.8\linewidth]{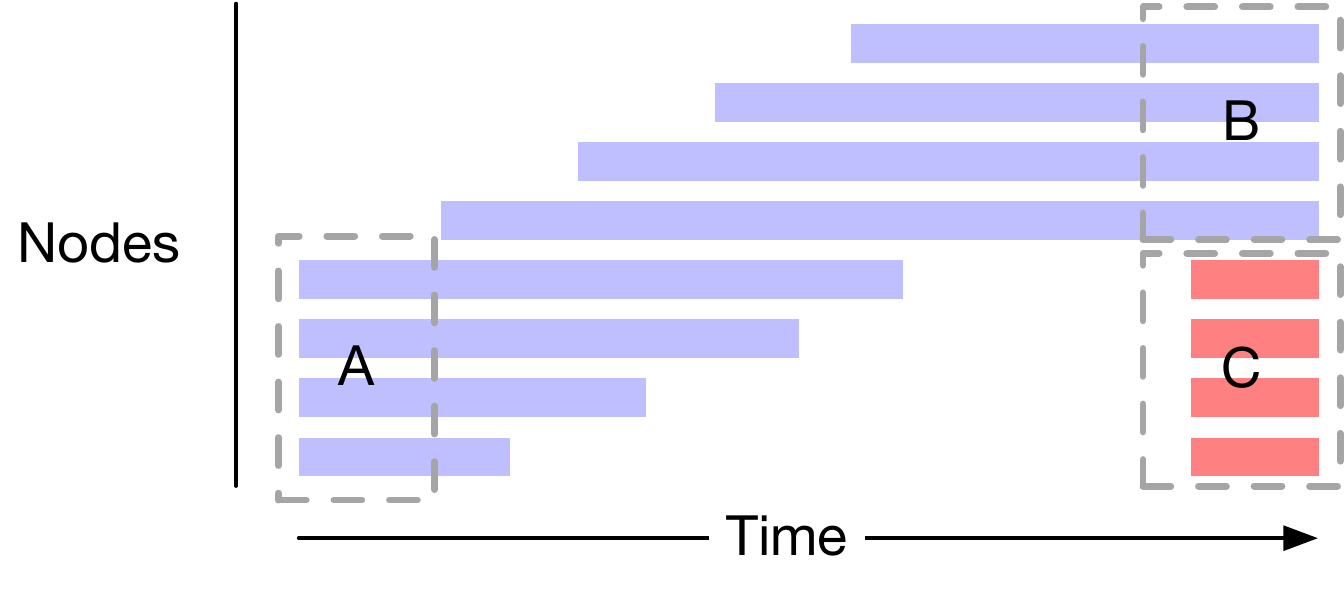}
    \caption{Illustration of the Ship of Theseus paradox. C is composed of the same nodes as A, but A progressively transforms into B. The choice to consider that B or C is \textit{the same community} as A therefore depends only on the community life-cycle.}
  \label{fig:ship}
\end{figure}

In terms of dynamic community detection, this scenario can be modeled (Fig. \ref{fig:ship}) by a progressively evolving community $c_1$(ship A), that nodes leave one after the others until all of them have been replaced (ship B). A new community $c_2$ appears after that, composed of the same nodes as the original community $c_1$ (ship C).

A static algorithm analyzing the state of the network at every step would be able to discover that there is, at each step, one community (at the beginning) and two at the end. But the whole point of dynamic community detection is to yield a longitudinal description, and therefore, to decide when two ships at different points in time are \textit{the same} or not.

The benchmark we propose is designed to represent complex evolution scenarios such as the ship of Theseus, in addition to usually defined events such as merge and split. It allows representing progressive changes for each event, e.g., add and remove edges incrementally such that two originally distinct communities merge into a single one. Unlike previous benchmarks, any community evaluation scenario can be described and generated using an appropriate language, and the network has both stable links and a community structure with known properties. We detail the difference with previous methods in section \ref{related}.

The rest of the paper is organized as follows. In section \ref{related}, we introduce previously proposed benchmarks, and we emphasize the added value of our proposition. In section \ref{benchmark}, we detail the generation process of our benchmark. Finally, in section \ref{experiments}, we compare several algorithms with different smoothing approaches on networks generated using the proposed benchmark.

\section{Related works}
\label{related}

A few methods have already been introduced in the literature to generate benchmark graphs for progressively evolving communities.

% Greene et al. \cite{greene2010tracking}, in one of the earliest works, proposed to generate sequences of static graphs using a widespread static generator (the LFR benchmark), by modifying some aspects of the communities at each step to create community events. This simple approach has drawbacks: $1)$ Edges are picked independently between snapshots, $2)$ Events are not progressive, which oversimplify the task of community tracking and event detection , and $3)$ Each event occurring has to be manually introduced by the experimenter.

%More recent benchmarks have tried to propose solutions to these drawbacks:

In Granell et al. \cite{granell2015benchmark}, two cyclic scenarios are proposed: one generates nodes migration (a set of nodes switch from a larger community to a smaller, and back), the other generates sequences of Merge-Split. In both cases, communities are defined as Stochastic Block Models (SBM), parameterized by fixed internal density $p^{in}$ and external density $p^{out}$.
%Events are generated as follows:
%\begin{itemize}
%\item For node migration, nodes switch one at a time, and for a switching node, all its edges are redrawn randomly to comply with its new affiliation.
%\item For Merge-Split, the process is defined by slowly rising/shrinking the number of inter-community edges until the community structure is well defined.
%\end{itemize}

In Bazzi et al. \cite{bazzi2016generative}, a generic method is introduced to generate multilayer networks with community structures. It requires to define an \textit{interlayer dependency tensor} encoding the probability for node $u_i$ in layer $l_a$ ($u_i,l_a$) to copy its community assignment from node $u_j$ in layer $l_b$ ($u_j,l_b$). In the most simple case, for dynamic networks, the community of ($u_i,l_t$) is defined as depending only on the affiliation of the same node in the previous layer ($u_i,l_{t-1}$). A random iterative process is used to attribute nodes to communities in each layer, satisfying both the constraints of the interlayer dependency tensor and a chosen distribution of community sizes. Edges are added in a second step, independently for each $t$, according to a degree corrected SBM, parameterized by a unique mixing parameter $\mu \in [0,1]$, where 0 corresponds to all edges falling inside communities, and 1 corresponds to no community structure. Note that edges are picked independently at each step, and events such are split or merge cannot be represented.
%Evaluation: Average NMI on each snapshot. GT: no problem
%Test With: Mucha with different parameters, multilayer InfoMap different param {Identifying modular flows on multilayer networks reveals highly overlapping organization in interconnected systems}

%In Greene et al. \cite{Greene}, probability at each step of having a node modification, add/remove, balance with a parameter for global growth. Edges are immediately added or removed to keep $p_{in}$ and $p_{out}$. Community operation can take place at any step. $p_{in}$ are assigned at random for split merge, (or inherited). A random process modify one edge at each step. This edge is selected randomly, with probabilities weighted to favor pairs of node with the highest difference between expected and observed density (problem of sub-structure...)
%Evaluation: -
%GT: superposed community structures
%Test with: -

In RDYN \cite{Rossetti2017}, community's lifecycles are created randomly based on rules: events (merge/split) have a probability to occur, and when an event occurs, involved communities are chosen by a random selection biased by the size of communities. Edges evolve gradually by combining two processes: a decay that makes old edges disappear, and a biased growth that reinforces community structure. The generation is driven by parameters, such as number of nodes, average degree, mixing coefficient, probability of node appearance, probability of node action, etc. Because the structure of communities at any given time is not necessarily well defined, the quality of communities is evaluated at each step using \textit{conductance}, and communities are included in the ground truth only when the value is higher than a threshold. This benchmark generates progressive events with complex lifecycles, but the evolution of communities is fully determined by internal mechanisms, and one cannot represent custom scenarios. The properties of generated communities are also not fully known, apart from their conductance property, since it is the result of an ad-hoc dynamic process.

%Evaluation: A custom quality function is defined. GT: determined by a quality function.

%Test with: iLCD, D-GT, TILES.

%Test With: Mucha with different parameters.

%In \cite{pasta2018model} <=== pas dynamique, processus dynamique...

The method proposed by Sengupta et al. \cite{sengupta2017benchmark} yields overlapping communities.
At each step, a community event (birth, death, split, merge) occurs with probability $p$ and a node event (add, remove) with probability $p'$. The initial community structure is generated by a static algorithm \cite{chykhradze2014distributed}. Edges inside each community are distributed randomly (Erdős-Renyi graph), with a probability decreasing with size: $\frac{\alpha}{n^\gamma}$ with $n$ the number of nodes and $\alpha,\gamma \in ]0,1[ $ chosen parameters. Edges are later modified randomly according to two factors: 1)random modifications at each step and 2) gradual evolution to match changes in the community structure. For instance, if two communities merge, edges to add are drawn at random among disconnected pairs of nodes in the resulting community in order to reach the desired density. Rules are introduced to ensure that communities stay coherent (Not becoming too small, not allowing simultaneous operations on the same nodes, etc.) Limitations are comparable to those of RDYN.

%Evaluation: Average weighted F1 (weighted by communities). Ground truth corresponds to assignment, no details on how to handle between change

%Tested on: OSLOM and MOSES

In several other articles, notably \cite{greene2010tracking,lin2008facetnet,tantipathananandh2011finding,ghasemian2016detectability, sarzynska2015null, zhang2017random, xu2014dynamic, benyahia2016dancer}, ad-hoc benchmarks were introduced, usually to test one specific scenario, with similar or more restricted scopes.
%(check also the Greene2010 paper)

Unlike all previous methods, the benchmark we propose introduces a language to represent any scenario of community evolution by specifying events (merge, split, etc.), either through its complete description, or by drawing randomly sequences of events (see section \ref{scenar} for examples). It also generates a network with 1) stable links (links present in $t$ are likely to be present in $t+1$), 2) communities with known properties (see section \ref{DSABM} for details), 3)Able to represent progressive events, such as a progressive merge or split.

To the best of our knowledge, a single paper has been published so far comparing empirically dynamic community detection algorithms: in \cite{coppens2019comparative}, 5 methods have been tested on RDyn benchmark \cite{Rossetti2017}.  They were compared in terms of average community quality at each step. In this article, we compare on different aspects, by introducing measures of smoothness and longitudinal quality (see Section \ref{evaluation})

% but none of them is able to create complex community evolutions such as the \textit{Theseus ship}(section \ref{theseus}) phenomenon.

\section{Synthetic network generation process}

\begin{figure}
    \centering
    \includegraphics[width=0.9\linewidth]{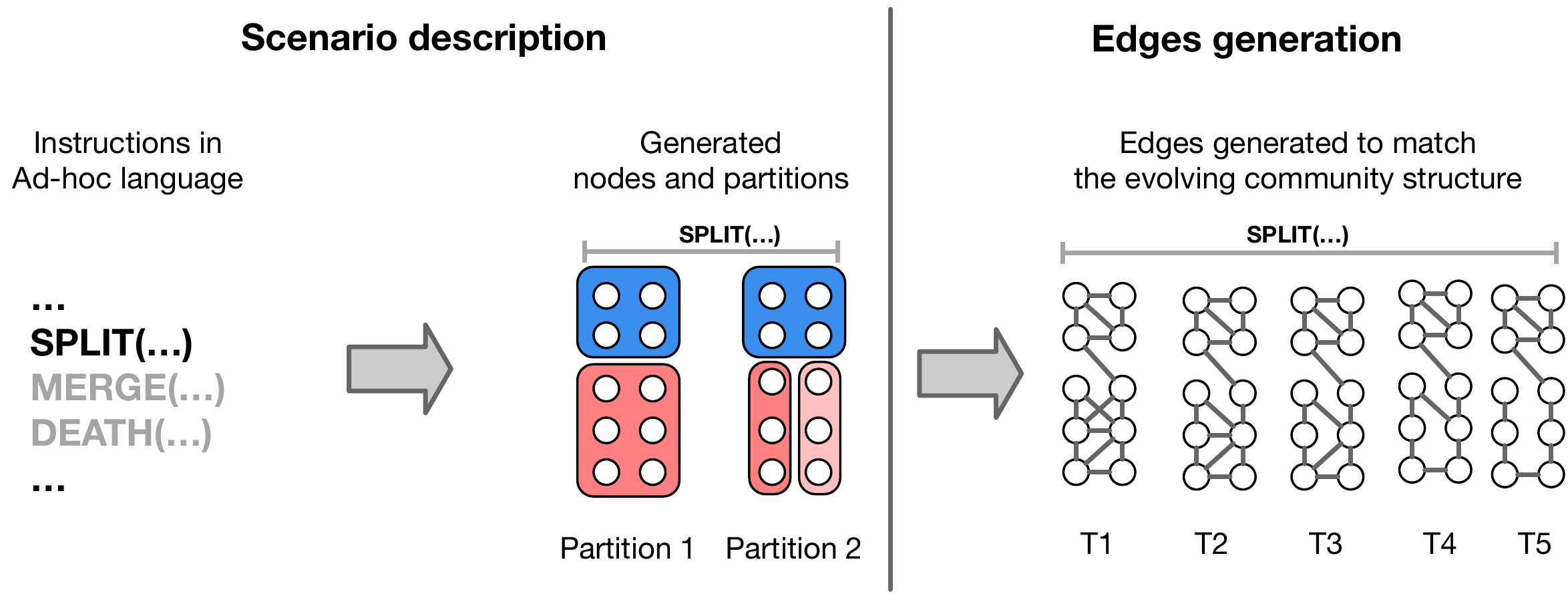}
    \caption{Illustration of the benchmark generation process.}
    \label{fig:schem}
\end{figure}

\label{benchmark}
The benchmark we propose follows a two-step process (Fig. \ref{fig:schem}):
\begin{enumerate}
	\item \textbf{Scenario description}: the experimenter defines initial communities and the scenario of their evolution (sequence of events).
	%\item \textbf{Assignment}: the affiliation of each node to its community is determined, for each time step
	\item \textbf{Edges generation}: edges are generated by a partly-random process. They form a dynamic network corresponding to the described community structure, satisfying some community quality properties.
\end{enumerate}

%\subsection{Definitions}
%\begin{Definition}
%	A community $c$ at time $t$ is defined as a couple $<N,l>$, with $N$ a set of nodes and $l$ a label.
%\end{Definition}
%\begin{Definition}
%	A partition $C_t$ at time $t$ is defined as a set of communities: $C_t=\{c_1,c_2,...,c_n\}$, all these communities being defined at time $t$.
%\end{Definition}
%\begin{Definition}
%	A dynamic community is defined a set of communities in different steps sharing a common label $l$
%\end{Definition}

%\makenomenclature
%\printnomenclature

\subsection{Scenario description}
Any scenario can be described by a set of \textit{community events}, each of them modifying the affiliation of nodes. To represent these community events, we define instructions allowing to represent the most common ones (Merge, Split, Birth, Death, etc.), or any arbitrary, more complex event.
%The evolution of communities is characterized by sequences of \textbf{Community events}.

%We define a set of \textbf{atomic events}, that can be combined to define \textbf{composed events}. Those events can be chained into more complex \textbf{event sequences}. A mechanism of \textbf{triggers} and \textbf{timers} allows to manage the order and timing of events.

%\subsubsection{Instructions definition}
We therefore define a language allowing to describe algorithmically any community evolution scenario. This language is composed of instructions, in the following form:
%of the form:
\vspace{0.5cm}

%\begin{algorithm2e}[H]
\textbf{C}$\gets$ \textbf{EVENT}(\textbf{parameters})[\textbf{delay},\textbf{triggers}]
%\end{algorithm2e}

\vspace{0.5cm}

Where
\begin{itemize}
		\item \textbf{C} is a list of communities yielded by this event, a community being a tuple $<ID,L,N>$ with:
		\begin{itemize}
			\item \textit{$ID$} a unique identifier for this community. Each community yielded by an event %event
			has a new, unique ID, e.g., an event that takes a community and removes one of its node will yield a community with a new unique ID.
			\item \textit{$L$} the \textit{label} associated with this community. The label corresponds to the \textit{identity} of the community: e.g., after a split, we can choose to attribute new labels to both resulting communities, or to give the label of the split
			community to one of the resulting ones.
			\item \textit{$N$} the set of nodes composing this community.
		\end{itemize}
	\item \textbf{EVENT} is the type of event (e.g., MERGE, BIRTH, etc.)
	\item \textbf{parameters} differ for each event. They are, for instance, communities to merge, labels associated to the
	yielded communities, etc.
	\item \textbf{triggers} is the set of communities that must be ready (have been yielded by a previous event) %to trigger this event.
	for this event to be triggered %(see section \ref{Triggers})
	\item \textbf{delay} is a number of steps to wait before starting the event after the triggers conditions are fulfilled. %(see section \ref{Triggers})
\end{itemize}
To sum up, the instruction above means that the event \textit{EVENT} will start \textit{delay} steps after all communities in \textit{triggers} appeared. This trigger/delay mechanism allows to define complex relations between communities, for instance the division of a community being triggered by the apparition of another  --topologically unrelated-- community.

We first define a single event, \textit{Assignment}, allowing to represent any change between an arbitrary number of communities. %We then define common events such as Merge or Split based on it.

%\subsubsection{Assignment Event}
\vspace{0.5cm}

\textbf{ASSIGN(BEFORE-COM, AFTER-NODES, AFTER-LABELS )}
\label{Atomic}

\begin{itemize}
	\item $BEFORE$ is an ordered list of communities that will be modified
	\item $AFTER-NODES$ is an ordered list of sets of nodes, each set of nodes corresponding to a community yielded by this event
	\item $AFTER-LABELS$ is an ordered list of labels to attribute to yielded communities.

\end{itemize}

\begin{algorithm2e}[H]
\SetKw{KwIs}{is}
\SetKw{KwTo}{to}
\SetKw{Kwglobal}{Global}

\SetKw{where}{where}
\SetKw{length}{length}

\KwIn{BEFORE-COM, AFTER-NODES, AFTER-LABELS}
$\Kwglobal$ AC : set of currently active communities

\Begin{

\For{c $\in$ BEFORE-COM}{
	AC $\gets$ AC  $\setminus$ c
}
NEW-COMS $\gets$ [] \;
\For{i $\gets$ 0  $\KwTo$   $\length$(AFTER-LABELS)}{
	NEW-COMS[i] $\gets$ $<$NEW-C-ID(), AFTER-LABELS[i], AFTER-NODES[i]$>$ \;
	AC $\gets$ AC $\cup$ NEW-COMS[i] \;
}

\Return  NEW-COMS
}
\caption{instruction \textbf{ASSIGN}. \textbf{NEW-C-ID()} is a function that generate a new, unique community identifier.}
\end{algorithm2e}
%\nomenclature{AC}{Set of Active Communities}%
%\nomenclature{NEW-C-ID()}{Function returning a new unique Community ID}%
%\nomenclature{NEW-N-ID()}{Function returning a new unique Node ID}%

%\subsubsection{Common events}
\vspace{0.5cm}
Most articles in the literature agree on a set of commonly found events impacting communities, such as SPLIT and MERGE.

In appendix \ref{events_definition}, we define some of these common events, based on the ASSIGN event: BIRTH, DEATH, MERGE, SPLIT, INITIALIZE and THESEUS, which corresponds to the ship of Theseus paradox presented in the introduction. The benchmark python library contains definition for additional events: CONTINUE (a community continues without change for a given period), RESURGENCE (a community disappears and re-appears with identical nodes some time later), and operations of progressive, node by node change: GROW-ITERATIVE, SHRINK-ITERATIVE and MIGRATE-ITERATIVE (nodes migrate from one community to another).

These instructions can be combined to define any scenario, either by listing all desired events or by writing a program to generate scenarios by picking instructions randomly. See section \ref{scenar} for examples.

%\textbf{GRADUAL-GROWTH(COM, NB-NODES, WAIT)}
%COM is the community to modify. NB-NODES correspond to the number of nodes to add, one after the other. WAIT is the time to wait between each node addition. The label of the provided community is conserved.
%
%We describe this composed event as a sequence of simple events.
%
%\noindent
%\underline{Algorithm}
%
%\begin{algorithm2e}[H]
%\SetKw{KwTo}{to}
%\SetKw{KwIs}{is}
%\SetKw{randomChoice}{randomChoice}
%
%
%\KwIn{COM, NB-NODES, WAIT}
%\SetKw{where}{where}
%
%\Begin{
%  \For{i $\gets$ 0 $\KwTo$ NB-NODES}{
%  		C $\gets$ BIRTH(1)[triggers=COM] \;
%   		COM $\gets$ MERGE([COM, C],[COM.L])[delay=WAIT,triggers=C])
%   	}
%  \Return COM
%}
%  \end{algorithm2e}
%

%\subsubsection{Description of scenarios}

%\subsection{Assignment}
%For several operations, it is convenient to avoid specifying exactly which nodes are affected. For instance, in a migrate operation, we can specify the \textit{number} of nodes that should
%The assignment step is straightforward from the Scenario definition. Note that several networks generated from the same Scenario might nevertheless lead to different assignments, leading to different lifecycles for nodes (but not for communities).

\subsection{Edges generation}
In the previous section, we have seen how to define the community evolution scenario. In this section, we address the generation of edges to fit this scenario.

Communities can have, at each step, two states: \textit{stable}, when it is not involved in any event, or \textit{evolving} otherwise. In the state \textit{stable}, edges of the community are generated following the \textit{Deterministic Strongly Assortative Block Model} (DSABM) (See Section \ref{DSABM}).

When an event is triggered, the communities involved switch to the \textit{evolving} state. Internal edges are known according to the DSABM before and after the event. Therefore, we make one edge modification (addition/removal) at each step of the dynamic network evolution, until reaching the final state.
At each step, external edges are also generated according to the DSABM.

%The edge generation process is performed independently for each step as follows:
%%To generate a dynamic network corresponding to the described community structure evolution, we adopt a chronological approach: starting from the initial state of the network, we check the list of all events ready to be performed. For each event, we do the following process:
%\begin{enumerate}
%	\item Observe the current state of the subgraph defined by nodes in input of the event
%	\item Compute the subgraph corresponding to the communities described in output of the event, according to a )
%	\item Compute the list of edges changes to go from the subgraph in input to the subgraph in outpout.
%	\item Perform one modification by graph evolution step to go from the input to the output subgraph.
%\end{enumerate}
%
%The edges between nodes belonging to different communities are computed at every step according to the DSABM.
%For a community not involved in any event, its internal edges are picked according to the DSABM.

%At each step, a static network is generated
%After the Scenario Definition step, the affiliations that each node must take is known. In the Edge generation step, we need to solve two intertwined problems:
%\begin{enumerate}
%	\item How to distribute edges so that the network topology corresponds to the defined scenario
%	\item How to make edges evolve to go from one stable community structure to another one
%\end{enumerate}

\subsubsection{Deterministic Strongly Assortative Block Model}
\label{DSABM}
%Deciding how the network should be built to correspond to the defined community structure is a crucial step since it depends on the \textit{definition} of community structure we adopt.
%Some graph generators define the community structure at the level of each node, e.g., the LFR benchmark \cite{} for which each node must respect a ratio between its intern and extern edges (Mixing coefficient). }
A common approach to generate static community benchmarks is to use a stochastic block model (SBM) \cite{holland1983stochastic}. %, in which the community structure is defined at the level of communities.
An edge probability matrix $P$ of size $r \times r$ is defined, with $r$ the number of blocks (communities), and edges between each pair of blocks are generated according to this probability matrix. We adopt a variation of the SBM, that we call \textit{Deterministic Strongly Assortative Block Model} (DSABM).

A block model is said to be \textbf{Strongly Assortative} if $P_{ii}>P_{ij}$ for each $i$ and $j$ such as $i \neq j$. This corresponds to the original definition of communities as \textit{groups of nodes that are more densely connected than the rest of the network}. Note that we adopted a Strongly Assortative structure for the sake of simplicity, the extension to other Block Models is straightforward.

We adopt a \textbf{deterministic} block model to solve the problem of the generation of \textit{slowly evolving} community structures. In the situation when one wants the network topology to follow a block structure and to allow this block structure to evolve, one needs to comply with two apparently antagonistic requirements:
\begin{itemize}
	\item \textbf{Economy of change}: we want as few edge modifications as possible to go from a network satisfying the partition $t_1$ to one satisfying another partition $t_2$.
	\item \textbf{Random internal structure}: we want the graph at each step to be compatible with the definition of block models, i.e. edges between communities should be chosen at random, and not depend on previous partitions.
\end{itemize}

These two requirements are antagonistic as soon as the internal density of communities changes between two steps. We can illustrate this problem with the following example: let's assume that we have at time $t$ two disconnected communities of size 4 and density 1. Each community contains 6 edges. Let's now consider that the two communities merge at time $t+1$ into a single community of density 0.5. This community should have 14 internal edges. If we try to maximize the \textit{economy of change} objective, we will add only two edges chosen randomly among the missing edges. The resulting community will thus be composed of two cliques of size 4 connected by only two edges, a structure very unlikely to obtain through a random edge selection for a single community of size 8 and density 0.5.

Conversely, maximizing the random internal structure by resampling independently networks at each step will lead to unstable edges, in particular for sparse blocks. This would not be compatible with a scenario of a progressive evolution from a network with a partition to a network with another one.

In previous benchmarks using SBM, the economy of change is usually ignored and random edges are generated at every step \cite{bazzi2016generative}.
In Granell et al.\cite{granell2015benchmark}, a solution is introduced: all pairs of nodes in the initial graph are ordered in a random fashion, and when edges need to be added or removed to reach the final state, they are chosen according to this predefined order
%ordered.

%This problem has been encountered by previous works on benchmarks for dynamic communities using SBM, with different approaches:
%\begin{enumerate}
%	\item In Bazzi et al. \cite{bazzi2016generative}, edges are sampled independently at each step, maximizing the Random Internal Structure.
	%\item In Greene et al. \cite{greene2010tracking}, a network is generated using an SBM for the first step. When the community structures evolve, edges are added (densification) or removed (sparsification) to comply with the new community structures, maximizing the Economy of change. Random edges modifications are introduced at each step to compensate for the bias introduced by previous partitions.
%	\item
%\end{enumerate}

We propose to generalize the method introduced in \cite{granell2015benchmark} to make it work for any type of scenario.
\vspace{0.5cm}

To each pair of nodes is assigned a fixed Latent Affinity score $\Omega \in [0,1]$. When a new node $n$ is added to a network $G=(V,E)$, a random value is assigned to each pair between $n$ and every node in $V$. When we need to attribute edges for a pair of blocks (communities $c_1,c_2$), we first compute the number $q$ of edges according to the probability matrix $P_{c_1 c_2}$ (Therefore interpreted as the \textit{fraction of existing edges} rather than an independent probability of observing each edge). We then select the $q$ pairs of highest $\Omega$ among pairs of nodes in this community.

The number and the position of edges in a block are therefore selected in a \textit{deterministic} way for a given community structure and %a
given latent affinities.
  Note that several runs of the same dynamic community scenario nevertheless lead to different dynamic networks, since latent affinities are assigned randomly.

\subsubsection{Communities density}
\begin{figure}[h!]
  \centering
  \begin{subfigure}[b]{0.45\linewidth}
    \includegraphics[width=\linewidth]{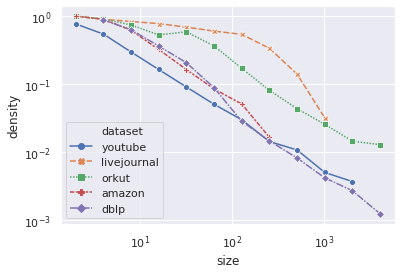}
    \caption{Average of densities of communities, \\ Ground Truth Communities}
  \end{subfigure}
  \begin{subfigure}[b]{0.45\linewidth}
    \includegraphics[width=\linewidth]{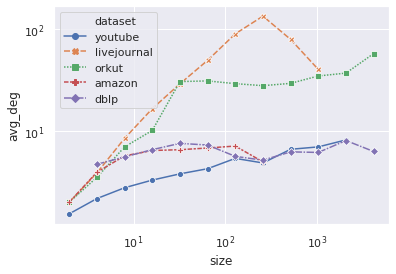}
    \caption{Average of internal mean degree of communities, communities discovered by the Louvain algorithm}
  \end{subfigure}

    \begin{subfigure}[b]{0.45\linewidth}
    \includegraphics[width=\linewidth]{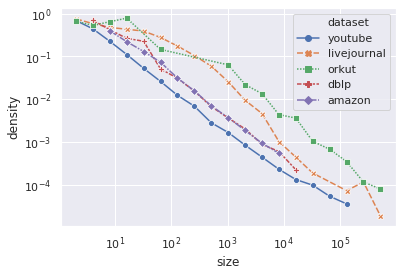}
    \caption{Average of densities of communities, \\ communities discovered by the Louvain algorithm}
  \end{subfigure}
  \begin{subfigure}[b]{0.45\linewidth}
    \includegraphics[width=\linewidth]{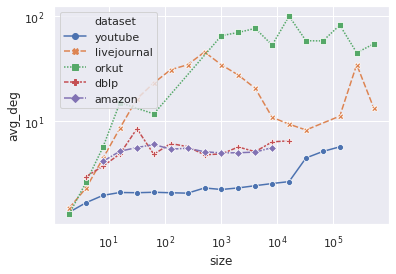}
    \caption{Average of internal mean degree of communities, communities discovered by the Louvain algorithm}
  \end{subfigure}

  \caption{relation between densities/mean degrees of communities relatively to their size in several large networks with available ground-truth communities. We observe similar trends for most networks: density decreasing with size, average internal degree increasing with size.}
  \label{fig:density}
\end{figure}

The DSABM used to generate the network corresponding to the desired community structure requires to define a Probability Matrix $P$. As for any Strongly Assortative SBM, we want $ P_{ii}>P_{ij}|i\neq j$. This is commonly solved by selecting two parameters $p^{ext}$ and $p^{in}$, such that $P_{ij}=p^{ext}$ and $P_{ii}=p^{in}$
%As for most Strongly Assortative models, we define an extern probability $p^{ext}$ such that $P_{ij}=p^{ext}$ for $i \neq j$. A common choice for $P_{ii}$ is to define a constant value $p^{in}$ and define $P_{ii}=p^{in}$.
However, we think that this choice is unrealistic for partitions with communities of heterogeneous sizes: intuitively, a community of size 3-5 must have an intern density $>0.7$ to be well defined, while a community of size 100 with a comparable density seems unlikely in empirical networks. A community \textit{growing} in size should therefore see its density \textit{shrink}. Additionally, it has been observed for dynamic graphs that the average degree tends to grow with the graph size\cite{leskovec2005graphs}. To validate empirically those intuitions, we compute the density and average degree of communities in a collection of large graphs with ground-truth communities from the SNAP dataset repository \cite{snapnets}. We use two definitions of communities:
%This phenomenon of shrinking density has already been observed for dynamic graphs \cite{leskovec2005graphs}. To confirm this property for communities, we computed the average density of communities in several large networks from the snap dataset \cite{snapnets}, according to:
\begin{enumerate}
	\item The so-called \textit{Ground Truth Communities}, corresponding to collected labels of nodes. Note that the communities found are not defined by the topology of the network, but based on meta-data
	\item Communities found in the same networks by the Louvain algorithm \cite{blondel2008fast}. Communities are thus topologically defined and correspond to a high value of Modularity.
\end{enumerate}
 We can observe in fig. \ref{fig:density} that the average density of communities tends to shrink with their sizes, while their average degree tends to grow, independently of the definition of communities considered. The slope of these trends depends on the network.

We therefore propose to model the density of each community using a simple function, compatible with observations.
%This function has the following properties:
% \begin{itemize}
% 	\item The average degree of nodes in community increases with its size (positive monotonic relationship)
% 	\item The density of a community decreases with its size (opposite monotonic relationship)
% \end{itemize}
%The only parameter required on top of the Scenario itself is the \textit{density coefficient}. The weakness of SBM based generators is that one needs to specify the desired density of communities. This problem is particularly important in dynamic settings. Let's consider a community growing progressively from 4 to 1000 nodes. Obviously, the density of the community of 4 should be high (e.g. $>\frac{2}{3}$.) for the community to be topologically well defined. But the same community grown to 1000 nodes with a density of $\frac{2}{3}$ would probably seem unrealistically dense.
We define the average degree of a community $c$, relatively to its number of nodes $n_c$ and a density coefficient $\alpha \in ]0,1]$ as:
\[
\bar{k_c}=(n_c-1)^\alpha
\]
The density of each community $c$ is thus defined as:
\[
p^{in}(c) = \frac{ n_c(n_c-1)^\alpha}{n_c(n_c-1)} =(n_c-1)^{\alpha-1}
\]
And the number of internal edges in a community:
\[
	m_c = \Bigl \lceil\frac{ n_c \times \bar{k_c} }{2}\Bigr \rceil
\]

%with $n_c$ the number of nodes in community $c$ and the density coefficient $\alpha \in ]0,1]$.
If $\alpha=1$, communities are cliques, and the density decreases faster with size for lower $\alpha$.

The external density $p^{ext}$ is defined as: $p^{ext}=\beta p^{in}(c_V)$ with $\beta$ a parameter of community identifiability and $c_V$ the whole graph seen as a community.

As a consequence, the internal and external density of all communities and their evolution with size are fixed with only two parameters: $\alpha,\beta\in[0,1]$.

These parameters can be chosen to vary the sharpness of communities as for any SBM based approach. In this article, we vary $\alpha \in [0.5,1]$ and $\beta \in [0,0.5]$.

\subsection{Random punctual noise}
Edges generated by the DSABM are deterministic for a given $\Omega$, i.e., if the communities do not change, the graph also stays unchanged. It has been proposed \cite{kobayashi2019structured} that in real dynamic networks, one can differentiate a stable \textit{backbone} from random, short-lived fluctuations. We add a parameter $\beta_r\in[0,1]$ to our benchmark, such that, at each step, a fraction $\beta_r$ of edges are rewired at random to differentiate the imperfections in community structures that are part of the backbone (controlled by $\beta$) and the ones which aren't ($\beta_r$)

%\subsection{Implementation}
%An implementation of the benchmark is available as part of a fully documented python package \footnote{\url{https://github.com/Yquetzal/tnetwork}}, and can be tested on-line as an interactive notebook \footnote{\url{https://colab.research.google.com/github/Yquetzal/tnetwork/blob/master/demo_generation.ipynb}}.

\subsubsection{Algorithmic complexity}
Each community being handled independently, the complexity of the generation process is not prohibitive: networks with hundreds of nodes and thousands of evolution step can be generated in a few seconds, and with thousands of nodes and tens of thousands of steps in a few minutes. The generation of very large networks is nevertheless not possible with the current implementation, due to the usage of the latent affinity score, which requires storing $n^2$ values. This constraint could be removed by using more advanced methods, for instance based on deterministic hash functions.

\begin{figure}
  \centering
  \begin{subfigure}[b]{0.90\linewidth}
    \includegraphics[width=\linewidth]{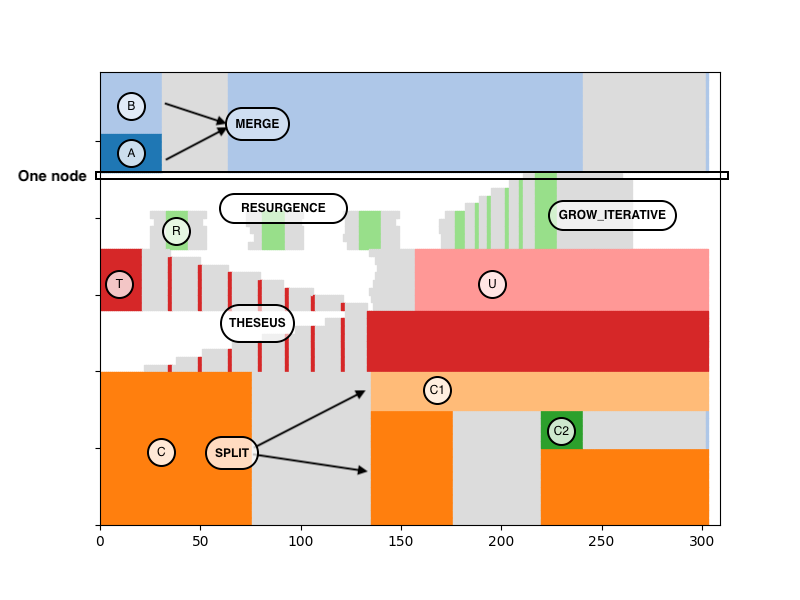}
    \caption{Planted dynamic communities, represented using the TAM visualization. Each node is represented as a thin horizontal line. Colors represent communities. Grey areas represent anbiguous affiliations. Events are identified by arrows and names, e.g., blue communities A and B merge into a singe community identified as B, and this process last from steps 30 to 60.}
  \end{subfigure}

  \begin{subfigure}[b]{0.45\linewidth}
    \includegraphics[width=\linewidth]{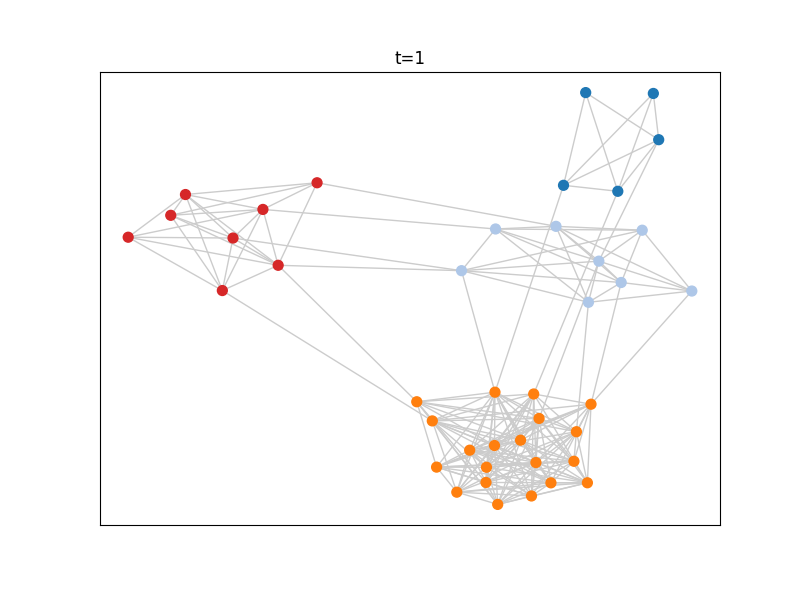}
    \caption{The static graph at time t=0, version \textit{sharp}\\ ($\alpha=0.9,\beta=0.05, \beta_r=0.01$)
    }
  \end{subfigure}
  \begin{subfigure}[b]{0.45\linewidth}
    \includegraphics[width=\linewidth]{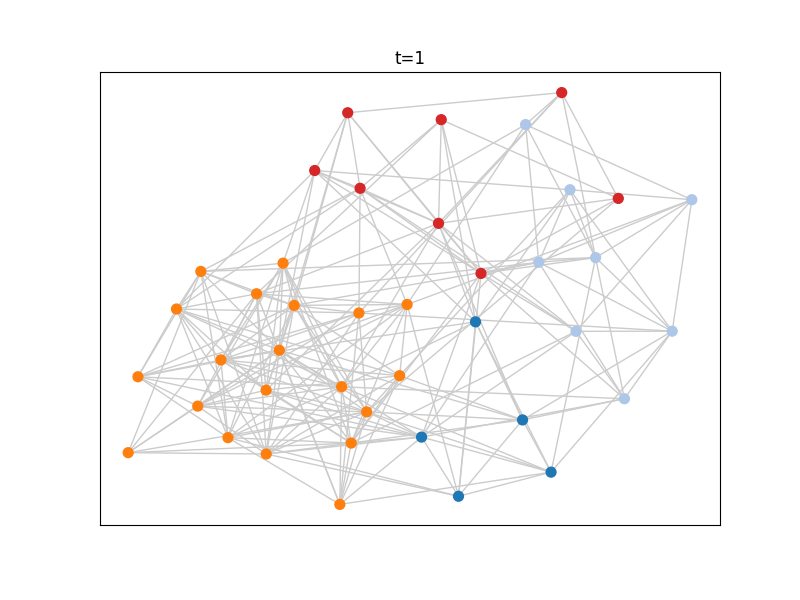}
    \caption{The static graph at time t=0, version \textit{blurred} \\($\alpha=0.8,\beta=0.25, \beta_r=0.01$ )}
  \end{subfigure}
  \caption{A simple scenario of community evolution. %In figure (a) we use
  }
  \label{fig:scenario}
\end{figure}

\begin{figure}
  \centering
  \begin{subfigure}[b]{0.45\linewidth}
    \includegraphics[width=\linewidth]{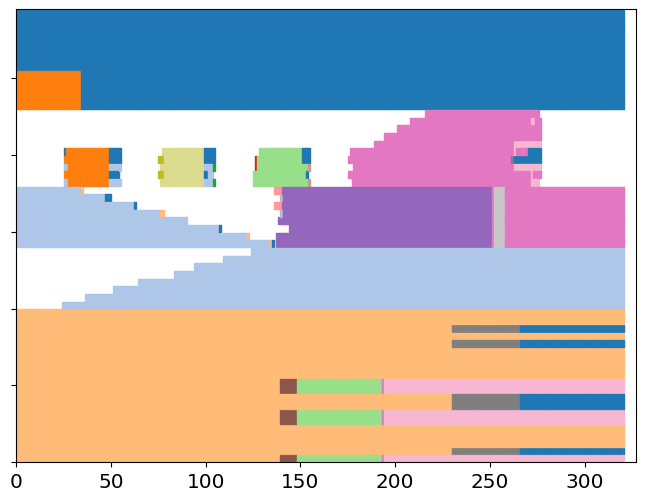}
    \caption{No-Smoothing}
  \end{subfigure}
  \begin{subfigure}[b]{0.45\linewidth}
    \includegraphics[width=\linewidth]{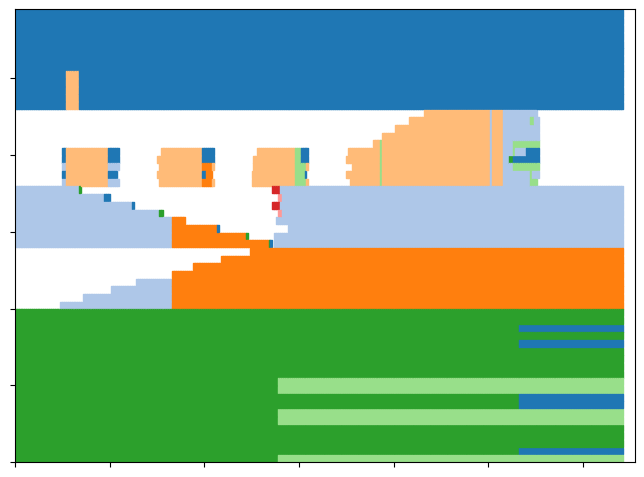}
    \caption{Label-Smoothing}
  \end{subfigure}
  \caption{Comparison of partitions obtained using two different methods on the ad-hoc scenario, \textit{sharp} flavor. Most communities are captured accurately, with some key differences: resurgence events are identified by Label-Smoothing but not by the other, The ship of theseus is labeled differently, etc.  }
  \label{fig:sharp}
\end{figure}

\begin{figure}
  \centering
  \begin{subfigure}[b]{0.45\linewidth}
    \includegraphics[width=\linewidth]{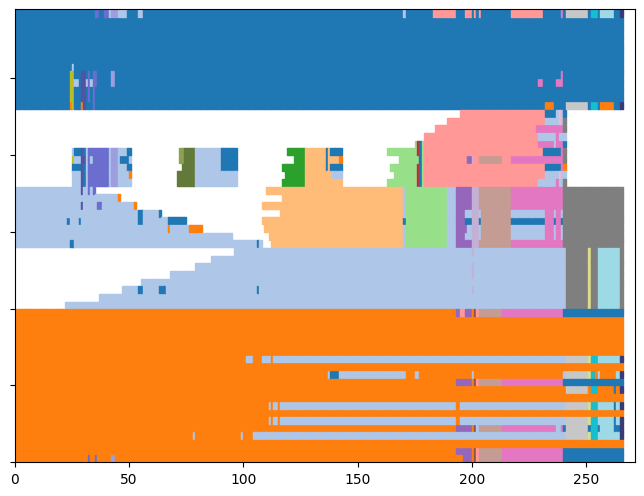}
    \caption{No-Smoothing}
  \end{subfigure}
  \begin{subfigure}[b]{0.45\linewidth}
    \includegraphics[width=\linewidth]{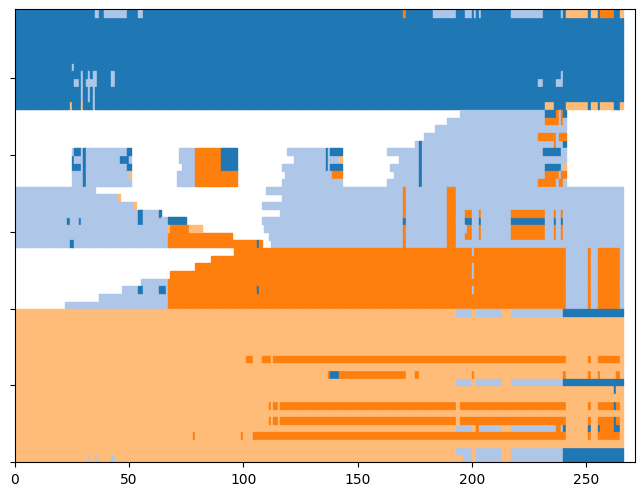}
    \caption{Label-Smoothing}
  \end{subfigure}

  \begin{subfigure}[b]{0.45\linewidth}
    \includegraphics[width=\linewidth]{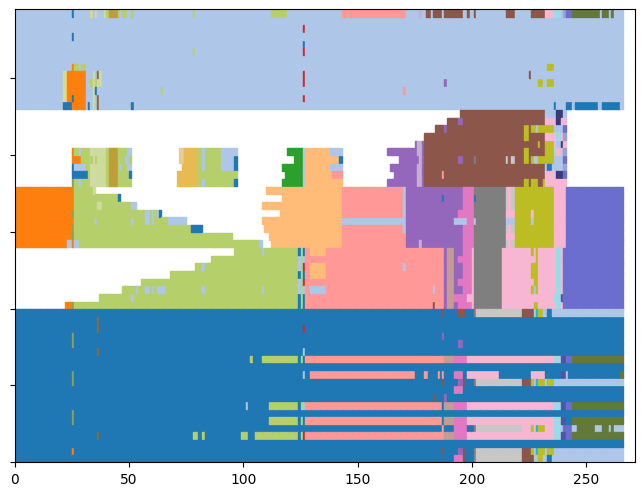}
    \caption{DYNAMO}
  \end{subfigure}
  \begin{subfigure}[b]{0.45\linewidth}
    \includegraphics[width=\linewidth]{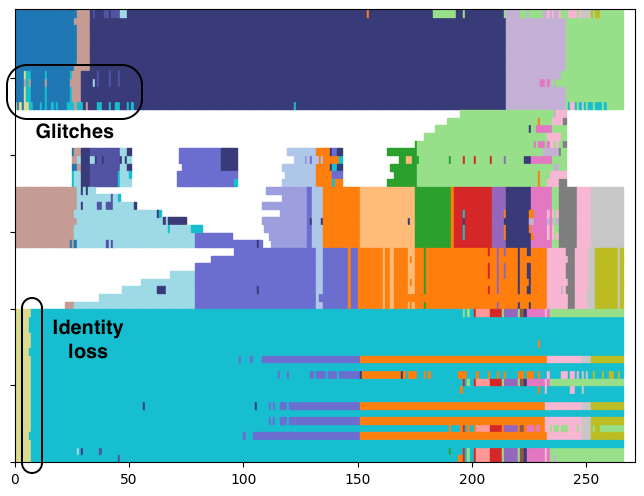}
    \caption{Transversal-Network}
  \end{subfigure}

  \begin{subfigure}[b]{0.45\linewidth}
    \includegraphics[width=\linewidth]{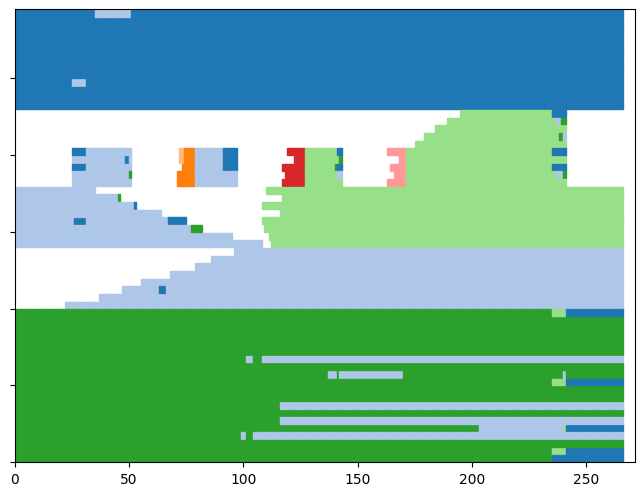}
    \caption{Implicit-Global}
  \end{subfigure}
  \begin{subfigure}[b]{0.45\linewidth}
    \includegraphics[width=\linewidth]{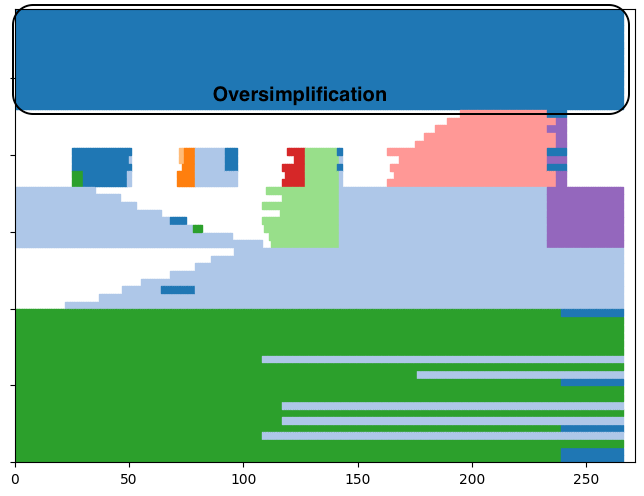}
    \caption{Smoothed-Graph}
  \end{subfigure}
  \caption{Comparison of partitions obtained using all methods on the ad-hoc scenario, \textit{blurred} variant. We annotated exampled of typical smoothing problems: Glitches, Identity loss and Oversimplification }
  \label{fig:blurred}
\end{figure}

\section{Experiments}
\label{experiments}
In this section, we evaluate several algorithms on networks generated using the proposed benchmark. We first introduce the algorithms to compare, and then conduct three experiments: qualitative evaluation on a complex scenario, quantitative evaluation on randomly generated networks, and scalability evaluation.

\subsection{Algorithms}
Many algorithms for Dynamic Community Detection have been proposed in recent years \cite{rossetti2018community}. Among them, we have selected six algorithms with different smoothing techniques. Our goal is not to search for the best algorithm, but rather to study the consequences of choices made to integrate the dynamic into the community detection process.
We have selected algorithms based on the following criteria:
\begin{itemize}
	\item Being based on \textit{Modularity optimisation}. We want all methods to agree on the definition of the best static partition on a single network, so that their differences depend only on the dynamic of the network. We chose the Modularity optimization approach because it is the most widespread, although a similar work could be done with SBM or Matrix factorization \cite{li2020optimization} based approaches, for instance.
	\item They represent well the variety of approaches used to tackle the dynamic aspect
	\item Their source code is available, or implementing them faithfully is not too difficult.
	\item They are scalable enough. We had for instance to discard popular methods such as DYNMOGA \cite{folino2013evolutionary}, Estrangement Confinement \cite{kawadia2012sequential} and FacetNet \cite{lin2008facetnet}, whose complexity is not compatible with having hundreds of steps of evolution.
\end{itemize}

The algorithms compared in this paper are the following:

\textbf{No-Smoothing}: The approach we will use as a reference consists in applying a static algorithm on the snapshot at each step, and then matching the most similar communities in consecutive steps, based on the Jaccard Coefficient. We use the Louvain method \cite{blondel2008fast} at each step, and the matching process, common to several approaches, is described in section \ref{label_method}.

\textbf{Implicit-Global} This method introduced in \cite{aynaud2010static} uses a form of \textit{implicit smoothing} \cite{rossetti2018community}: at each step, the Louvain algorithm is run, but instead of starting it with each node in its own community, the previous partition is used as seed.

\textbf{DYNAMO} \cite{zhuang2019dynamo} is a recent method updating at each evolution step the community structure according to changes in the graph, based on a set of local rules. The primary goal is to be faster than \textit{No-smoothing} while reaching similar Modularity scores, by avoiding to recompute communities from scratch at each step. However, as Implicit Global, it also introduces some smoothing by staying close to a previous local minimum (\textit{implicit local} smoothing). We used the implementation by the authors.
%Previous methods with the same mechanism exist \cite{x,y}, but DYNAMO claims to improve some of their weaknesses, and its code is available.

\textbf{Smoothed-Graph}
This method is a variant of the one proposed in \cite{guo2014evolutionary}. A community detection algorithm (in our case, Louvain) is run at each step $t$ on a graph whose smoothed adjacency matrix is defined as follows: $A^t_{ij}=\alpha A^{t}_{ij}+(1-\alpha)C^{t-1}_{ij}$ where $C^{t-1}_{ij}=1$ if $i$ and $j$ belongs to the same community at step $t-1$, $0$ otherwise.

%Explicity Trade-Off approaches ha
%\cite{folino2010multiobjective} is explicitly trying to smooth the community structure by optimizing at each step $t$ a multi-objective quality functions, considering the modularity at $t$ and the NMI with the partitions at $t-1$. It does so using a genetic algorithm.

\textbf{Transversal-Network} is a popular method introduced by Mucha et al. \cite{mucha2010community} with a \textit{Cross-Time} approach, i.e. communities at $t$ depends on earlier and later steps of the network. The principle of the method is to build a single transversal network by adding inter-snapshot coupling links and to apply a Louvain-like community detection algorithm on this network, based on an adapted version of the modularity. We used the original implementation by the authors. %For the sake of performance and simplicity, we used a more recent implementation in the Leidenalg package \cite{traag2009community}

\textbf{Label-Smoothing} is a method introduced by Falkowski et al. \cite{falkowski2007data} whose first step is the same as the No-Smoothing algorithm, but instead of matching communities between pairs of successive steps, a \textit{Community survival graph} is created by considering each community in each step as a node, and an edge connects any two communities with a Jaccard coefficient above a threshold, with the Jaccard value as weight. Community detection is applied to this graph to define \textit{communities of static communities}, thus defining dynamic communities. We implement it using the Louvain algorithm for both steps, and similar parameters as for the matching method described in section \ref{label_method}.

All algorithms have several parameters that could be modified to improve the results. However, community detection being an un-supervised problem by definition, these parameters usually cannot be tuned. We therefore used the default parameters from the implementation (DYNAMO) or used by the authors themselves ($\omega=0.5$, \cite{mucha2010community}). $\alpha=0.9$ \cite{guo2014evolutionary})

\subsubsection{Persistent labels attribution}

\hfill \break
\label{label_method}
Methods \textit{No-Smoothing}, \textit{Implicit-Global}, \textit{DYNAMO} and \textit{Smoothed-Graph} detect communities at each step, with or without smoothing, but do not attribute persistent labels to communities. For all those methods, we therefore attribute persistent labels to communities using the same procedure, inspired by \cite{greene2010tracking}. First, a similarity score is computed between any two pairs of communities between adjacent snapshots using the Jaccard coefficient, and any pair of communities with a value of similarity above a threshold (0.3, in our experiments) is considered a potential match. Then, two communities $c_a$ in $t$ and $c_b$ in $t+1$ are matched if i) $c_b$ is the most similar community to $c_a$ in $t+1$ AND ii) $c_a$ is the most similar community to $c_b$ in $t$. If a community in $t+1$ is not matched to any community in $t$, it receives a new label.
When creating our benchmark, we respect this logic, i.e., in case of merge or split, the communities the most similar before and after the event share the same label (we avoid ties in our scenarios).

\subsection{Qualitative evaluation on an ad-hoc scenario}
  \label{scenar}

In this section, we generate a small, deterministic scenario to observe qualitatively how each algorithm behaves. The scenario is designed to include several particular cases such as a Ship of Theseus, resurgence of communities, and successive merge and split of communities. It is described as follows (note that the algorithm is a functional python code, using the provided implementation library):

\vspace{0.15cm}
%\begin{minted}[xleftmargin=8pt, linenos, fontsize=\small]{python}
\begin{lstlisting}[language=Python,caption={Defining an ad-hoc scenario. Resurgence is an operation that makes a community disappear and reappear with the same label after a delay.},label={code:scenar}]

    # Initialization with 4 communities of different sizes
    [A, B, C, T] = my_scenario.INITIALIZE([5, 8, 20, 8],["A", "B", "C", "T"])

    # Create a theseus ship after 20 steps
    (T,U)=my_scenario.THESEUS(T, delay=20)

    # Merge two of the original communities after 30 steps
    B = my_scenario.MERGE([A, B], B.label(), delay=30)

    # Split a community of size 20 in 2 communities of size 15 and 5
    (C, C1) = my_scenario.SPLIT(C, ["C", "C1"], [15, 5], delay=75)

    # Split again the largest one, 40 steps after the end of the first split
    (C1, C2) = my_scenario.SPLIT(C, ["C", "C2"], [10, 5], delay=40)

    # Merge the smallest community created by the split, and the one created by the first merge
    my_scenario.MERGE([C2, B], B.label(), delay=20)

    # Make a new community appear with 5 nodes, disappear and reappear twice, grow by 5 nodes and disappear
    R = my_scenario.BIRTH(5,  name="R", delay=25)
    R = my_scenario.RESURGENCE(R, delay=10)
    R = my_scenario.RESURGENCE(R, delay=10)
    R = my_scenario.RESURGENCE(R, delay=10)

    # Make the resurgent community grow by 5 nodes 4 timesteps after being ready
    R = my_scenario.GROW_ITERATIVE(R, 5, delay=4)

    # Kill the community grown above, 10 steps after the end of the addition of the last node
    my_scenario.DEATH(R, delay=10)

\end{lstlisting}

We generate two networks using this scenario with different parameters, a \textit{sharp} scenario ($\alpha=0.9,\beta=0.05$), and a \textit{blurred} one ($\alpha=0.8,\beta=0.25$). We add a small random noise $\beta_r=0.01$.

In fig. \ref{fig:scenario}, we can see the ground truth communities corresponding to this scenario, together with the initial state of the network at $t=0$ for both sets of parameters.
To represent the planted communities, we use the Temporal Activity Map (TAM \cite{linhares2017dynetvis}) visualization approach, i.e., each node has a fixed vertical position, edges are not represented, time is on the horizontal axis, and colors correspond to community affiliations (two nodes with the same color belongs to the same community, whether they are in the same timestep or different ones). Nodes appear grey when they have no known affiliation, which corresponds to periods during which events are \textit{on-going}, affected communities not being properly defined. Nodes not present in the network at a given time appear white.

In fig. \ref{fig:sharp}, we can compare the results obtained by the No-Smoothing and the Label-Smoothing approaches on the sharp variant. We can observe that the non-smoothed algorithm already matches quite well the ground truth, without too much instability. When using the Label-Smoothing approach, despite having the exact same partition at each step initially, some important differences arise: i)the resurgent community is identified as such (yellow community for all of them), ii)The ship of Theseus is %splitted
split differently.%, communities composed of the same nodes having the same labels.

In fig. \ref{fig:blurred}, the same results are shown for all methods on the blurred variant. From this figure, we define three types of problems that can be observed in dynamic communities:
\begin{itemize}
    \item \textit{Glitches} corresponds to individual nodes switching arbitrarily between communities for short periods.
    \item \textit{Identity loss} corresponds to a community label being lost and replaced by a different one due to a short-lived change, while the community stayed mostly coherent.
    \item \textit{Oversimplification} corresponds to topologically distinct communities at a given time being merged to improve the smoothness of the solution.
\end{itemize}

We can observe that there are much fewer glitches in some smoothed approaches (Implicit-Gobal, Smoothed-Graph) compared with No-Smoothing or DYNAMO.
Similarly, Identity loss observed in the No-Smoothing approach for the ship of Theseus (orange $\rightarrow$ green $\rightarrow$ blue $\rightarrow$ purple, etc.) is partially solved by Label-Smoothing, Implicit-Global and Smoothed-Graph.
However, Oversimplification also appears, for instance with communities A and B being considered as a single one for Smoothed-Graph.

\begin{figure}
  \centering

   \begin{subfigure}[b]{0.3\linewidth}
    \includegraphics[width=\linewidth]{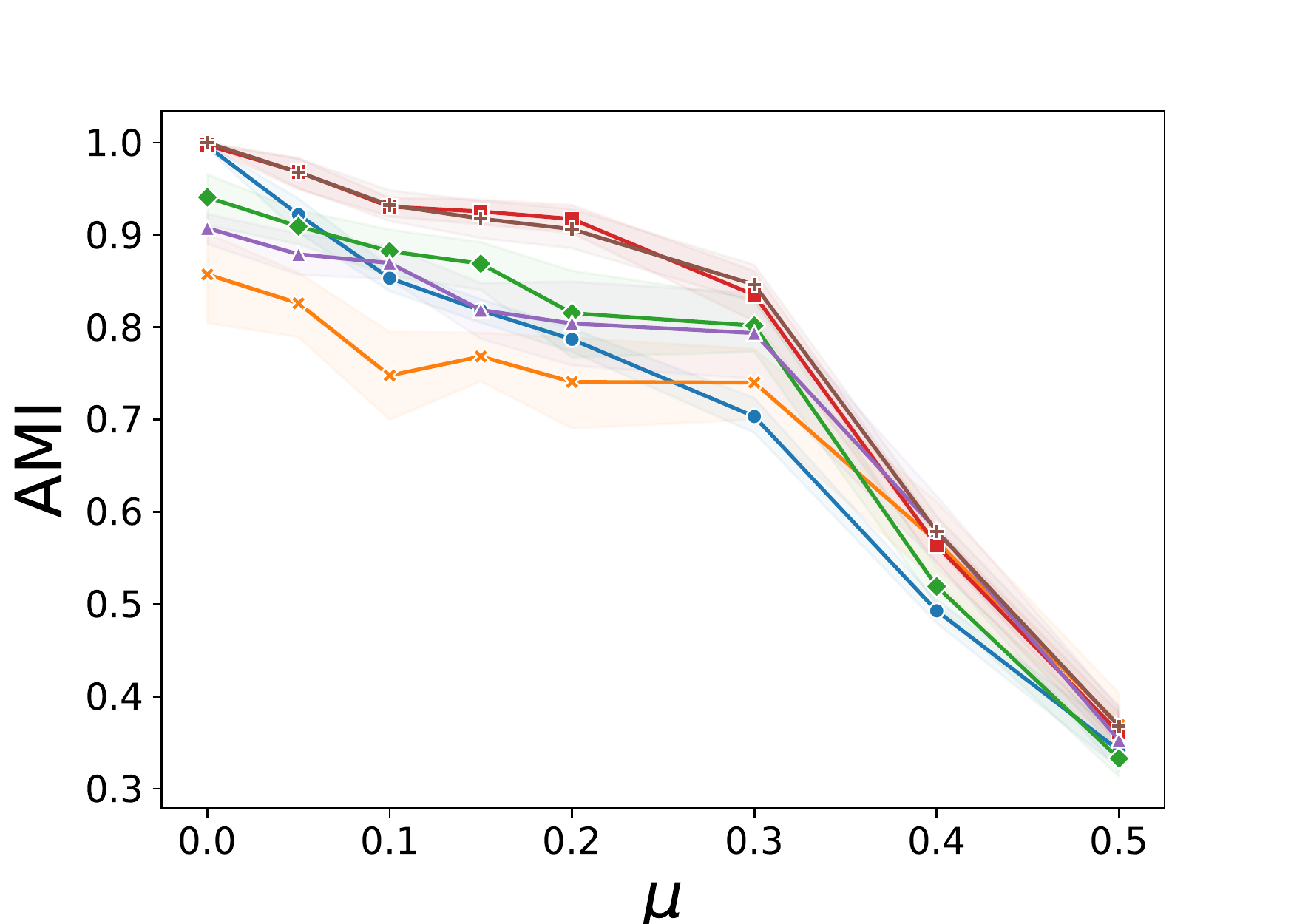}
    \caption{$\overline{AMI}$, average AMI}
  \end{subfigure}
  \begin{subfigure}[b]{0.3\linewidth}
    \includegraphics[width=\linewidth]{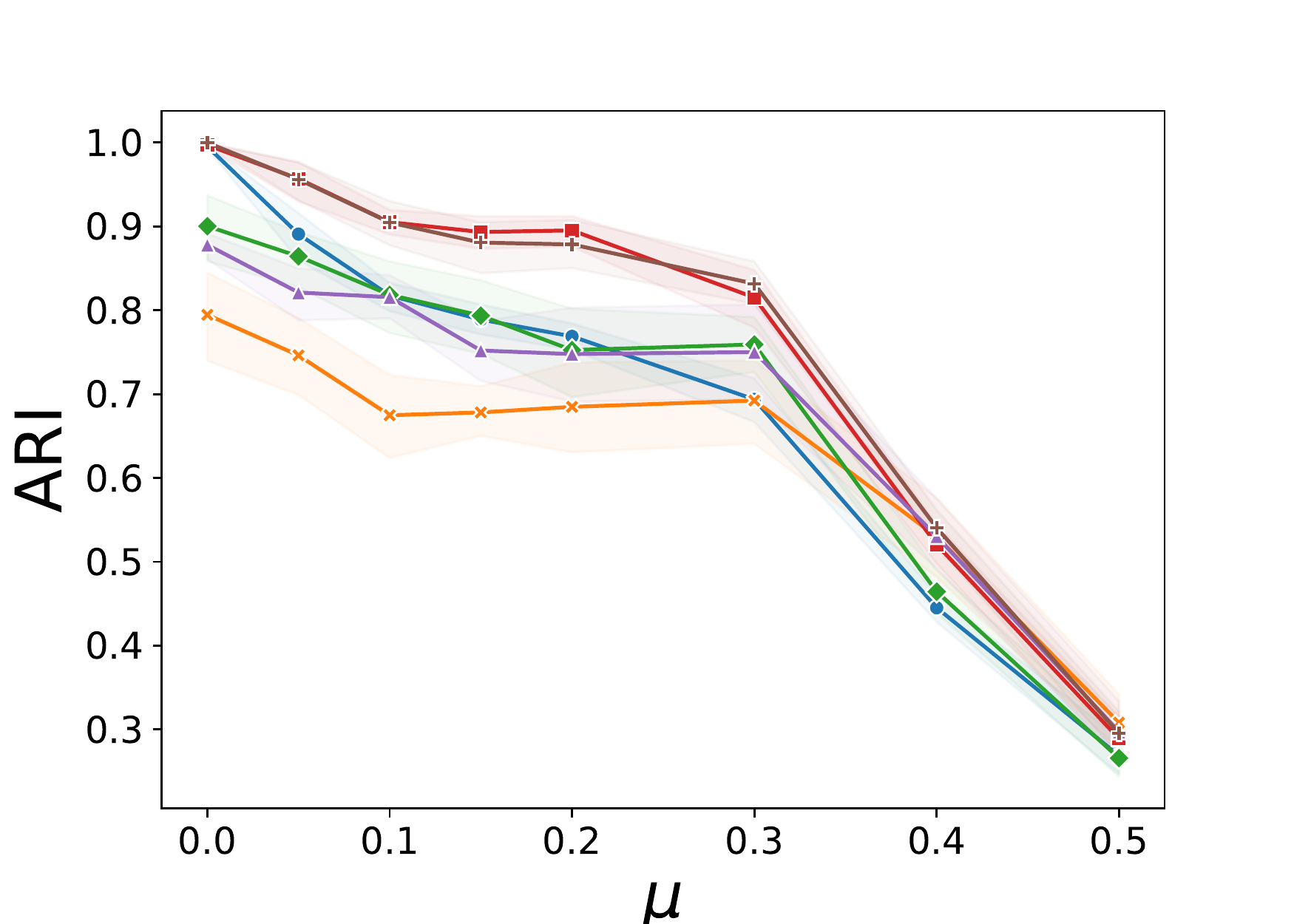}
    \caption{$\overline{ARI}$, average ARI}
  \end{subfigure}
   \begin{subfigure}[b]{0.3\linewidth}
    \includegraphics[width=\linewidth]{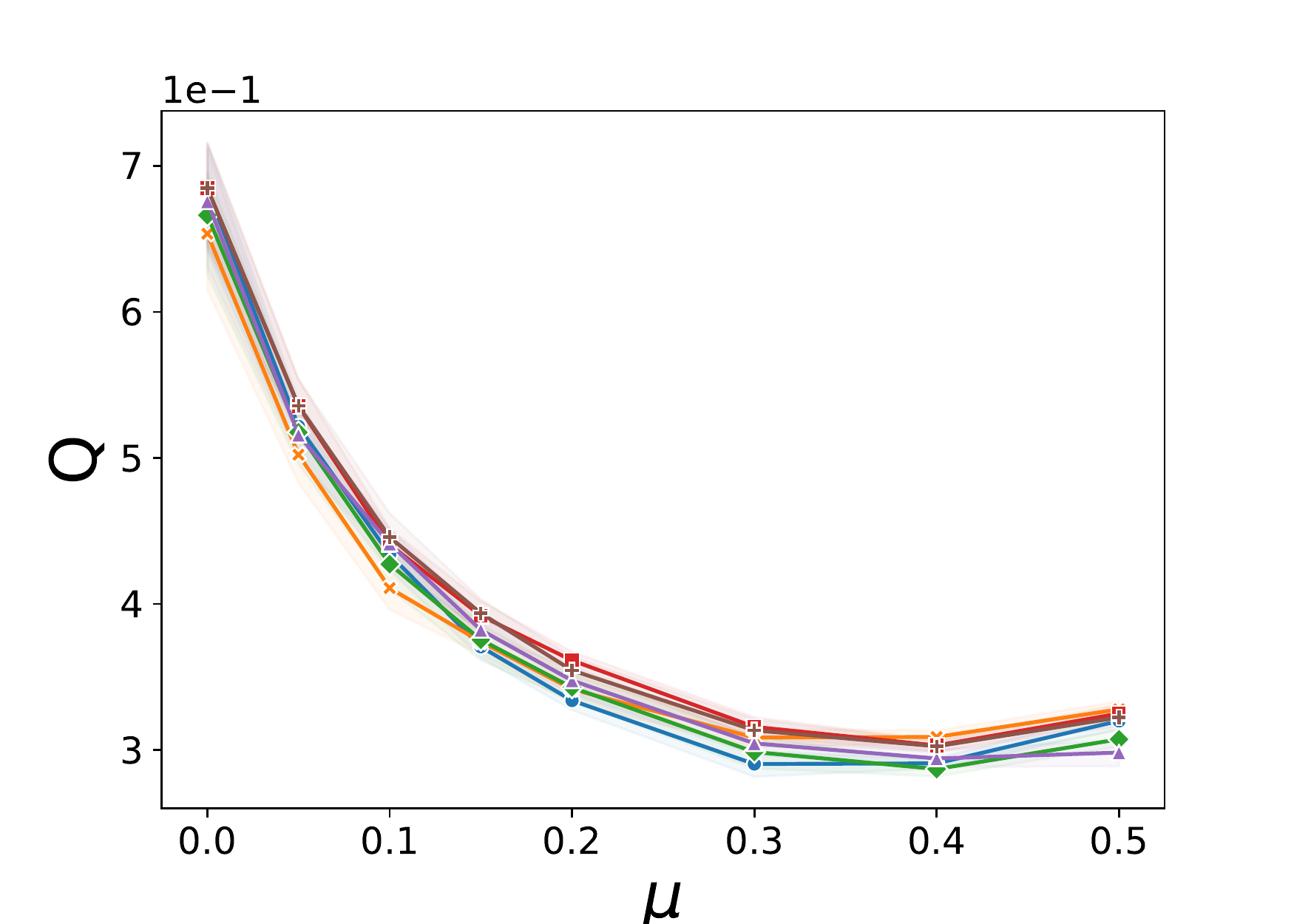}
    \caption{$\overline{Q}$, average Modularity}
      \end{subfigure}

	\begin{subfigure}[b]{0.3\linewidth}
    \includegraphics[width=\linewidth]{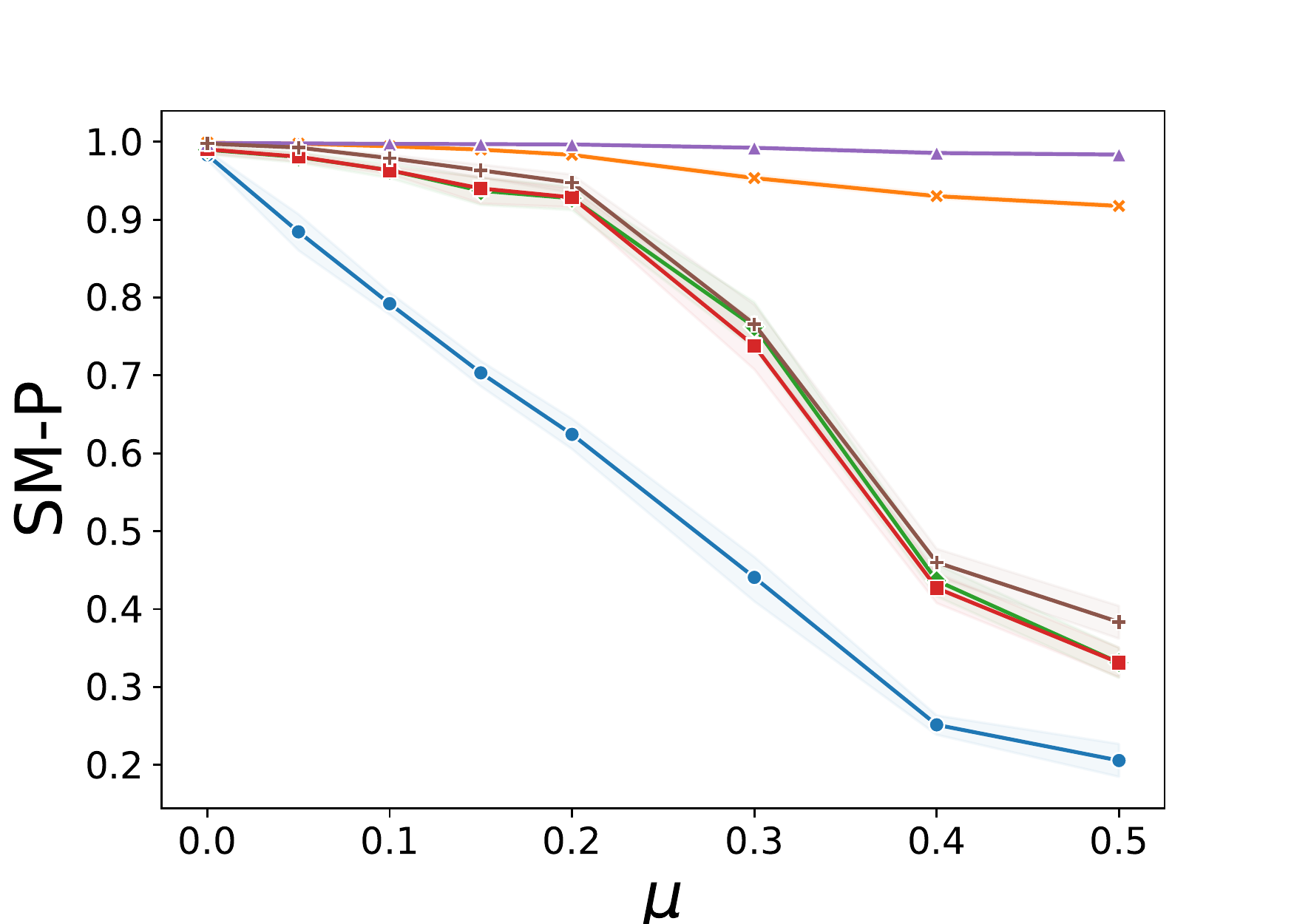}
    \caption{SM-P, Partition smoothness}
  \end{subfigure}
   \begin{subfigure}[b]{0.3\linewidth}
    \includegraphics[width=\linewidth]{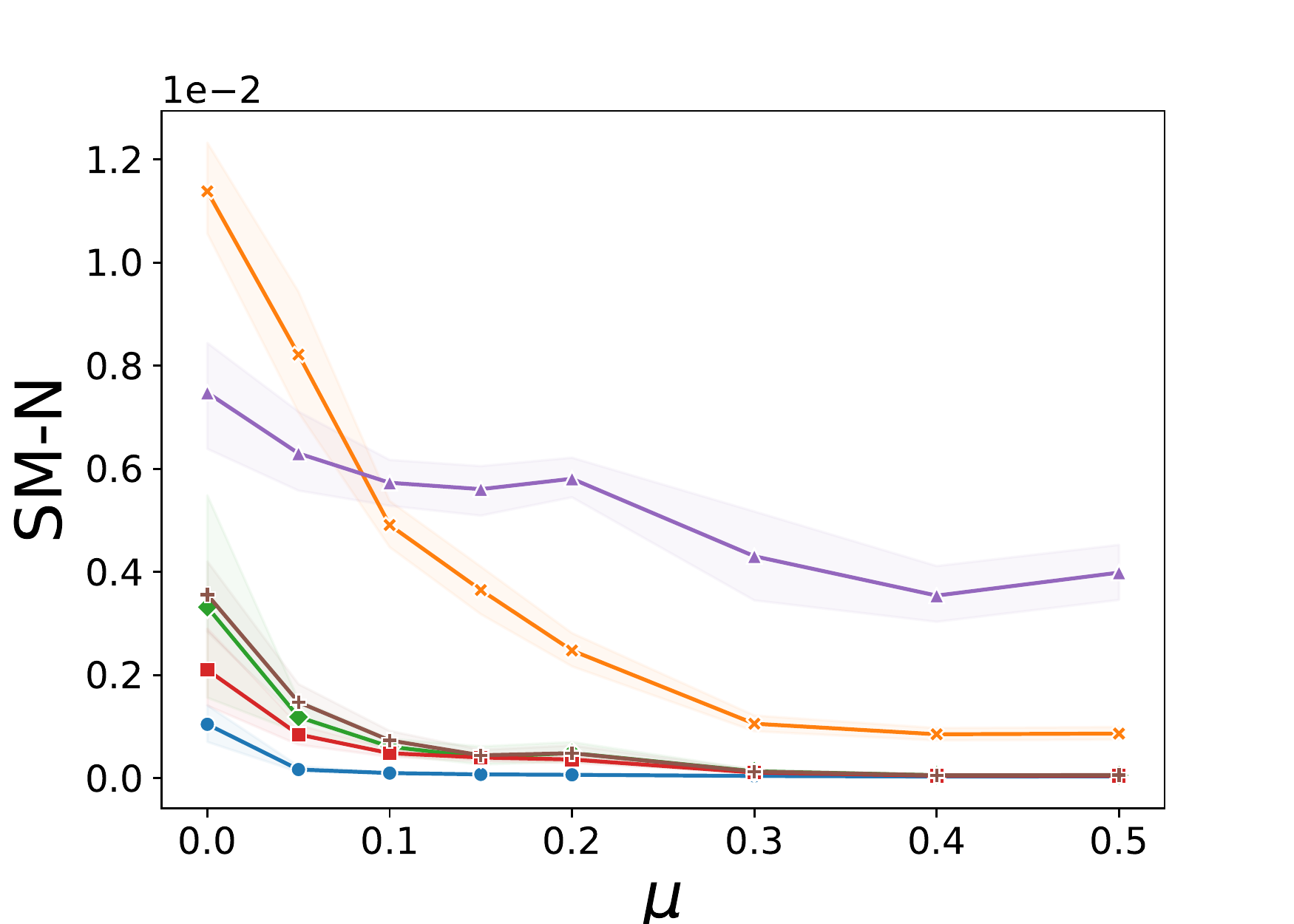}
    \caption{SM-N, Node smoothness}
  \end{subfigure}
  \begin{subfigure}[b]{0.3\linewidth}
    \includegraphics[width=\linewidth]{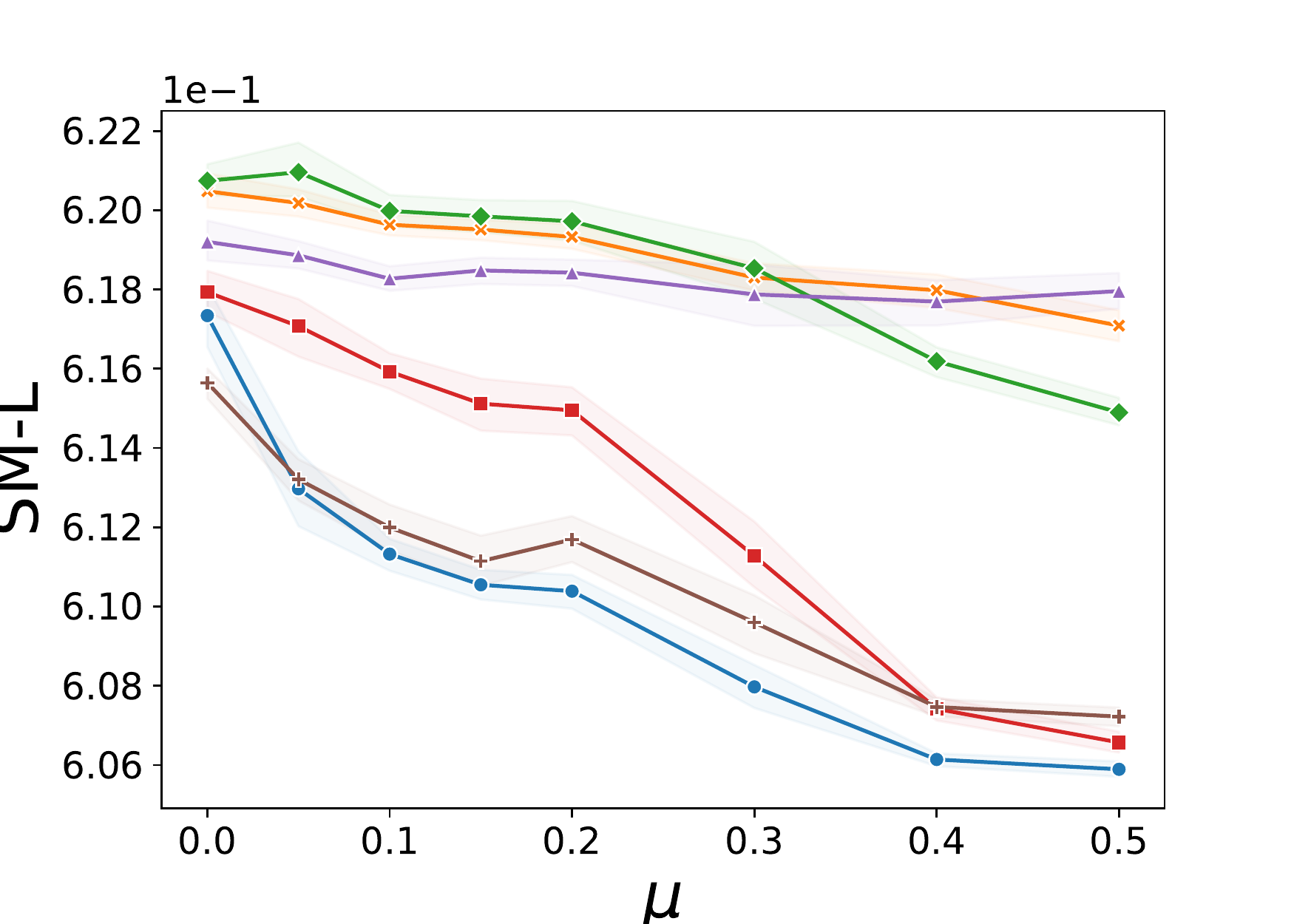}
    \caption{SM-L, Label Smoothness}
  \end{subfigure}

    \begin{subfigure}[b]{0.3\linewidth}
    \includegraphics[width=\linewidth]{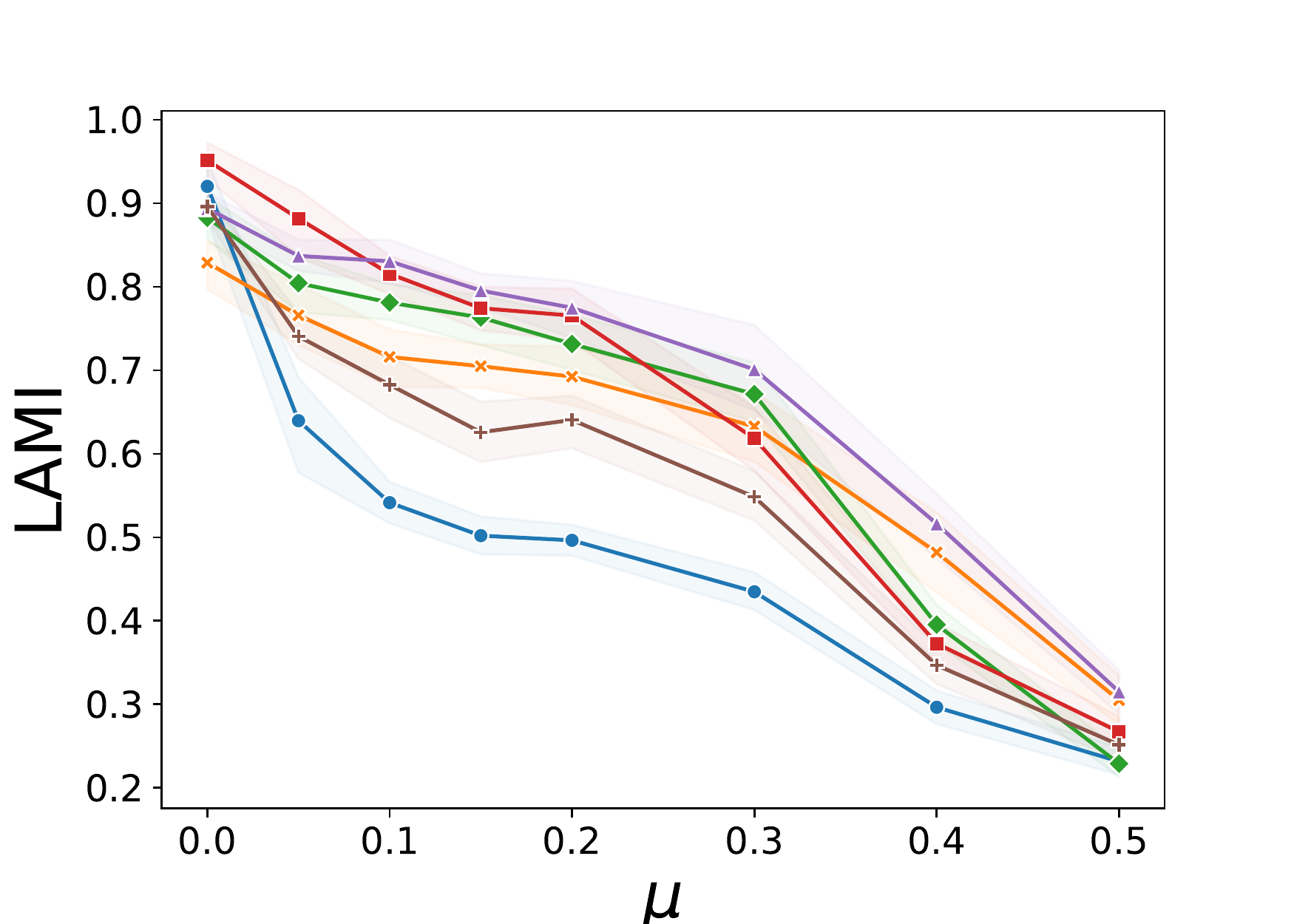}
    \caption{LAMI}
  \end{subfigure}
  \begin{subfigure}[b]{0.3\linewidth}
    \includegraphics[width=\linewidth]{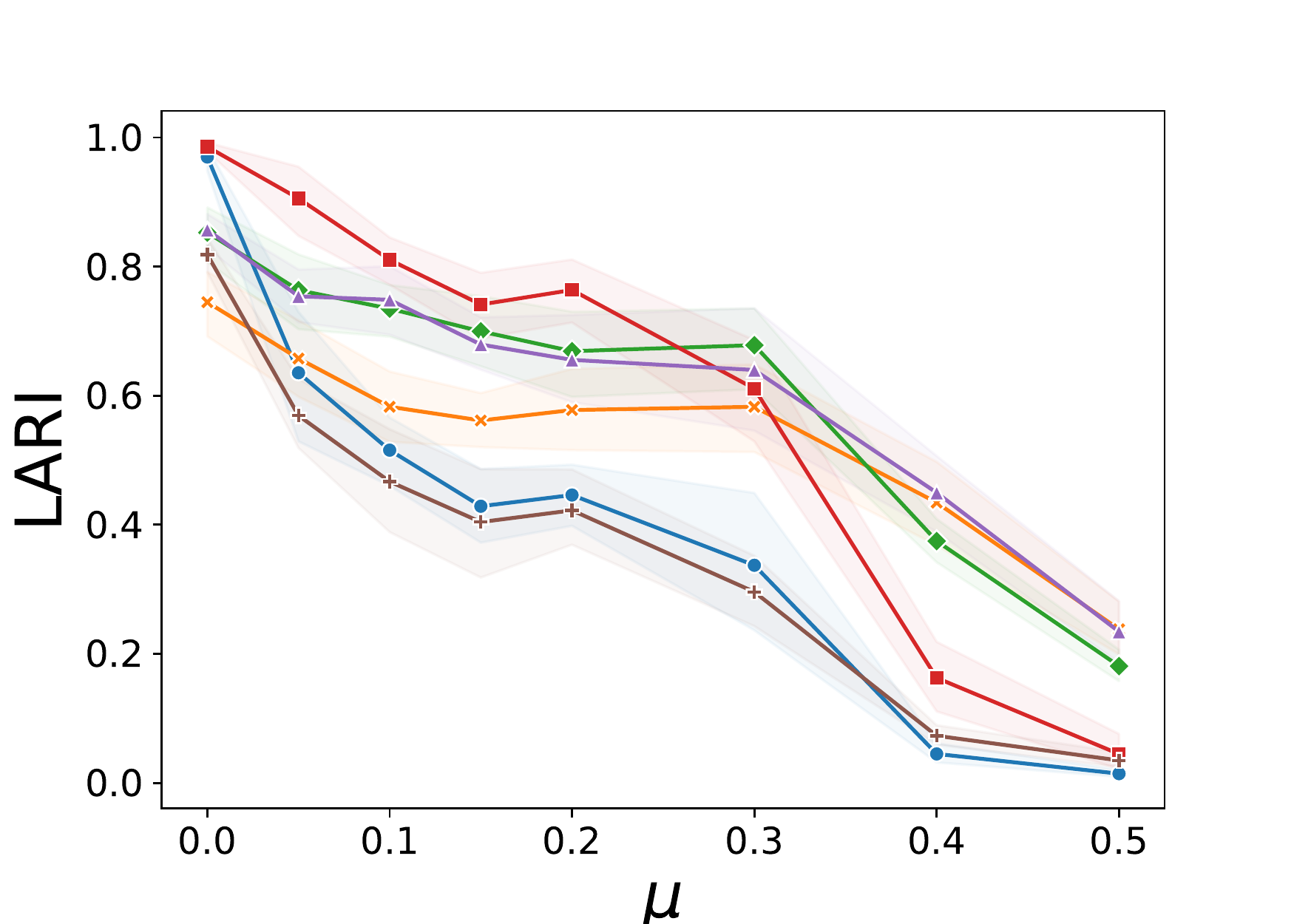}
    \caption{LARI}
  \end{subfigure}
    \begin{subfigure}[b]{0.3\linewidth}
    \includegraphics[width=\linewidth]{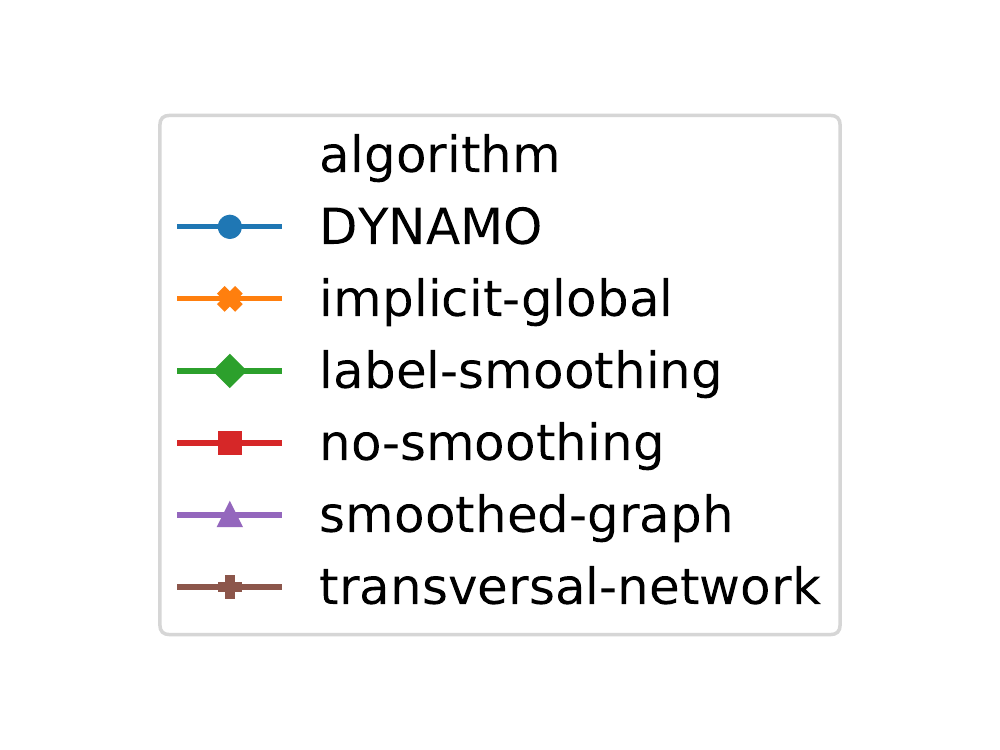}
    \caption{Legend}
  \end{subfigure}
  %\begin{subfigure}[b]{0.3\linewidth}
  %  \includegraphics[width=\linewidth]{LF1.pdf}
   % \caption{LF1}
 % \end{subfigure}

    \caption{Change in dynamic partitions properties when the $\mu$ parameter increases (higher score is better). Top row: Instantaneous Quality functions. Middle row: Smoothness quality functions. Bottom row: Longitudinal Quality functions. We observe clear differences between algorithms: no-smoothing and transversal network have among the highest scores in instantaneous Quality functions, but among the lowest in smoothness ones. Smoothed graphs and implicit global have the opposite behavior. Despite those opposite approaches, no-smoothing and smoothed graph both have high scores in longitudinal scores.}
  \label{fig:quality}
\end{figure}

\subsection{Quantitative evaluation of quality on a random scenario}
\label{evaluation}
In order to quantitatively evaluate the quality of dynamic communities discovered by each algorithm, we design a generator of random scenarios, based on the following principle: given a number of communities $m$, a minimal and maximal community sizes $s^{min}$ and $s^{max}$, and a number of operations $o$, we repeat $o$ times the following process (code available in the implementation):
\begin{enumerate}
	\item Pick a community $c$ at random.
	\item If $|c|>s^{max}$, split it in two, the largest resulting community inherits $\frac{2}{3}$ of its nodes.
	\item Else, merge it with the smallest remaining community.
\end{enumerate}

\subsubsection{Scores for the evaluation of dynamic communities}
All algorithms selected for evaluation have the same quality function in the static case: Modularity. We introduce quality functions to assess different aspects of the quality of dynamic communities: evaluation of their quality at each point in time, of their smoothness, and of their longitudinal quality.

\vspace{10pt}
\textbf{Evaluation at each step}

First, we use scores to evaluate the quality of communities at each step taken independently. As already done for instance in \cite{Rossetti2017}, we use the average values for each step of static scores. We use average Modularity $\overline{Q}$ to assess the intrinsic quality, average AMI $\overline{AMI}$ and average ARI score $\overline{ARI}$ to compare with the ground truth. AMI is the adjusted for chance version of the Normalised Mutual Information, while ARI is the adjusted rand index. Note that, in theory, a static approach run at each time without smoothing should have the highest values for these metrics, thus high scores in these metrics alone are not good measurements of the quality of dynamic communities.

\vspace{10pt}
\textbf{Evaluation of smoothness}

We introduce 3 scores to evaluate the smoothness of dynamic partitions.
\begin{itemize}
    \item \textbf{SM-P} is defined as 1 -
    the average NMI between all pairs of successive snapshots, and measure the \textit{smoothness at the level of \textbf{partitions}}. (Higher is better) \\ More formally, SM-P $= 1- \frac{1}{T-1}\sum_{t}^{T-1} NMI(G_t,G_{t+1}) $
    \item \textbf{SM-N} is defined as the inverse of the number of affiliation change (summed for all nodes). It measures the \textit{smoothness at the level of \textbf{nodes}}. (Higher is better) \\ More formally, SM-N = $1/(\sum_n^V \sum_t^{T-1} \delta_{L(n,t),L(n,t+1)}$), with $L(n,t)$ the label of node $n$ at time $t$ and $\delta_{i,j}$ the kronecker delta, i.e., $\delta_{i,j}=1$ if $i=j$, 0 otherwise.
    \item \textbf{SM-L} is defined as the inverse of the average Shannon entropy of node labels, i.e., for each node, we compute the entropy $H$ of its label affiliations, and SM-L is the inverse of the average among all nodes. It measures the \textit{smoothness at the level of \textbf{labels}}. (Higher is better) \\ More formally, SM-L $=1/ \sum_n^V H(L(n))$, with $H$ the Shannon Entropy and $L(n)$ the probabilities of observing each label when picking node $n$ at a random time $t$.
\end{itemize}

It is important to note that they measure different aspects of smoothness: SM-P is independent of labels and is little impacted by the instability of single nodes%,
. SM-N is sensible to glitches (short-lived, spurious changes)%, while
. SM-L is little sensible to glitches but is impacted by long-term instability, e.g., an unjustified label change which is not reversed.
\vspace{10pt}

\textbf{Longitudinal scores}

Finally, we introduce scores to compare dynamic partitions with a ground truth longitudinally.
Let's define the longitudinal partition as follows:
\begin{Definition} A \textbf{longitudinal partition} associates a label $l$ to each set of tuples $(n,t)$, with $n$ a node and $t$ a timestep.
\end{Definition}
Given a reference longitudinal partition $L^{ref}$ and a longitudinal partition to compare $L$, any static community comparison function can be applied on those longitudinal partitions as with any partition. We apply AMI and ARI comparison function, hereafter called LAMI and LARI.

\subsubsection{Experimental settings}
We fix parameters as $m=10$, $s^{min}=5$, $s^{max}=15$, $o=20$.
We define an initially sharp partition with $\alpha=1$ and $\beta=0$. We make the sharpness vary by using a parameter $\mu$ such as $\alpha=1-\mu$ and $\beta=\mu$. We also add random noise $\beta_r=0.01$. We repeat each experiment 20 times. On average, the resulting networks have 100 nodes and 1200 steps (snapshots).

\subsubsection{Results}

\begin{figure}
    \centering
    \includegraphics[width=0.8\linewidth]{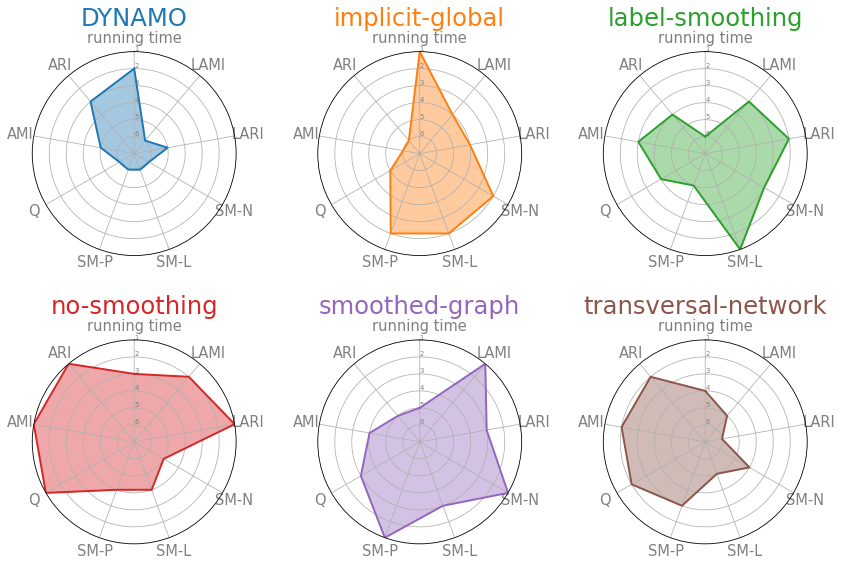}
    \caption{Radar chart of ranks of different methods for $\mu=0.2$. Higher score is better. We observe that the No-Smoothing method obtains among the best results in instantaneous scores (Q,AMI,ARI), while Smoothed-Graph and Implicit-Global obtain higher scores in smoothing scores. Label-Smoothing has the highest score in label smoothness.}
    \label{fig:spider}
\end{figure}

Fig.\ref{fig:quality} synthesizes the results by scoring method, while Fig.\ref{fig:spider} synthesizes the results by method, for a value of $\mu=0.2$, which seems to be a tipping point. We make the following observations:
\begin{itemize}
	\item In terms of instantaneous quality ($\overline{AMI}$, $\overline{ARI}$, $\overline{Q}$), as expected the No-Smoothing approach has the highest values, together with the Transversal-Network, for all $\mu$. We can note that $\overline{AMI}$ starts at 1 (for $\mu=0$) and do not fall below 0.3 for any method, showing that the community structure, although blurred, stays roughly detectable.
	\item In terms of smoothness, two methods have high scores for the three aspects: Implicit-Global and Smoothed-Graph. Label-Smoothing has the highest scores in most settings for the $SM-L$ scores, which measure label smoothness. DYNAMO is the least stable in most cases.
	\item No method seems to be a clear winner in terms of Longitudinal similarity with the ground truth. We can note however that for sharp communities ($\mu<0.15$), the No-Smoothing approach always obtains the highest score, while for higher $\mu$, Smoothed-Graph and in some cases Label-Smoothing obtain the highest scores. This is coherent with the observation that static algorithms become unstable when the community structure is not unambiguously defined, and that smoothing is therefore needed to obtain stable dynamic communities.
	%\item In terms of Longitudinal similarity to the ground truth, both scores mostly agree on the scale of relative scores, but not on the ranking. Transversal-Networks and DYNAMO have the lowest scores for both, while Label-Smoothing, No-Smoothing and Smoothed-Graph have the highest. When communities are very well defined, No-Smoothing obtains the highest scores in all cases, while, when noise increase, it is Smoothed-Graph (for LAMI) and label-smoothing (for LARI).
	\end{itemize}

%\textbf{Average of static scores}

From Fig.\ref{fig:spider} , we can make the following additional observations: Smoothed-Graph and Implicit-Global provide the strongest smoothing, but, as a consequence, compared with the reference No-Smoothing method, they have communities of lower quality in each individual snapshot.
%The DYNAMO approach yields results comparable to the No-Smoothing one, although slightly less smoothed.
%The Transversal-Network approach, although yielding instantaneous results comparable to those of the reference, seems to be notably less smoothed in term of label.

\subsection{Scalability evaluation}
Another important aspect to consider in comparing methods is their capacity to handle large networks with many steps of evolution. The complexity of the No-Smoothing approach, for instance,  is simple to estimate. It can be defined as $TO^{CD}+(T-1)O^{M}$, with $T$ the number of steps, $O^{CD}$ the complexity to run the static community detection algorithm at each step, $O^{M}$ the complexity of the matching process between consecutive partitions. The first part, which is the most costly, can be trivially parallelized. The computational complexity is therefore linearly proportional to the number of steps and depends on the size of the network at each step and the chosen static algorithm. Other algorithms have complexities that depend on other factors and are harder to formulate theoretically in comparable terms.

In this section, we compare empirically the scalability of the chosen methods by varying two parameters: either we fix the size of the network at each step, and vary the number of steps, or the contrary. More formally, we use the same algorithm to generate random dynamic graphs as before, but:
\begin{itemize}
    \item In the first test, we change the number of operations $o=50$ in order to have a large number of steps, and run computations on subsets of the first $x$ steps, varying $x$.
    \item In the second test, we vary the number of initial communities $m$ while keeping the same average size. We run algorithms on a slice composed of the first 50 steps only.
\end{itemize}

\begin{figure}
  \centering

  \begin{subfigure}[b]{0.4\linewidth}
    \includegraphics[width=\linewidth]{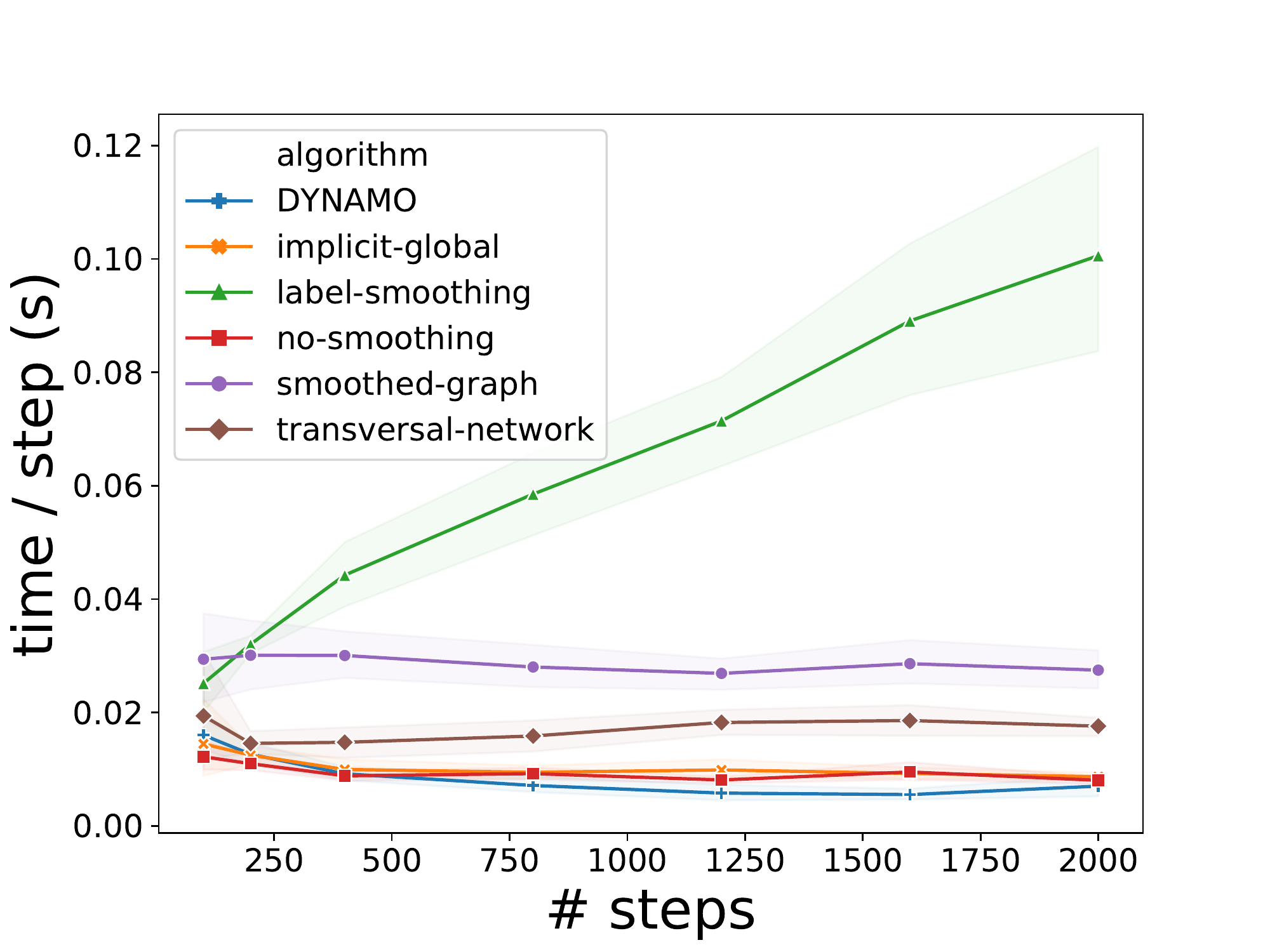}
    \caption{Scalability for different number of steps, for a same size of graphs (n=50)}
\end{subfigure}
   \begin{subfigure}[b]{0.4\linewidth}
    \includegraphics[width=\linewidth]{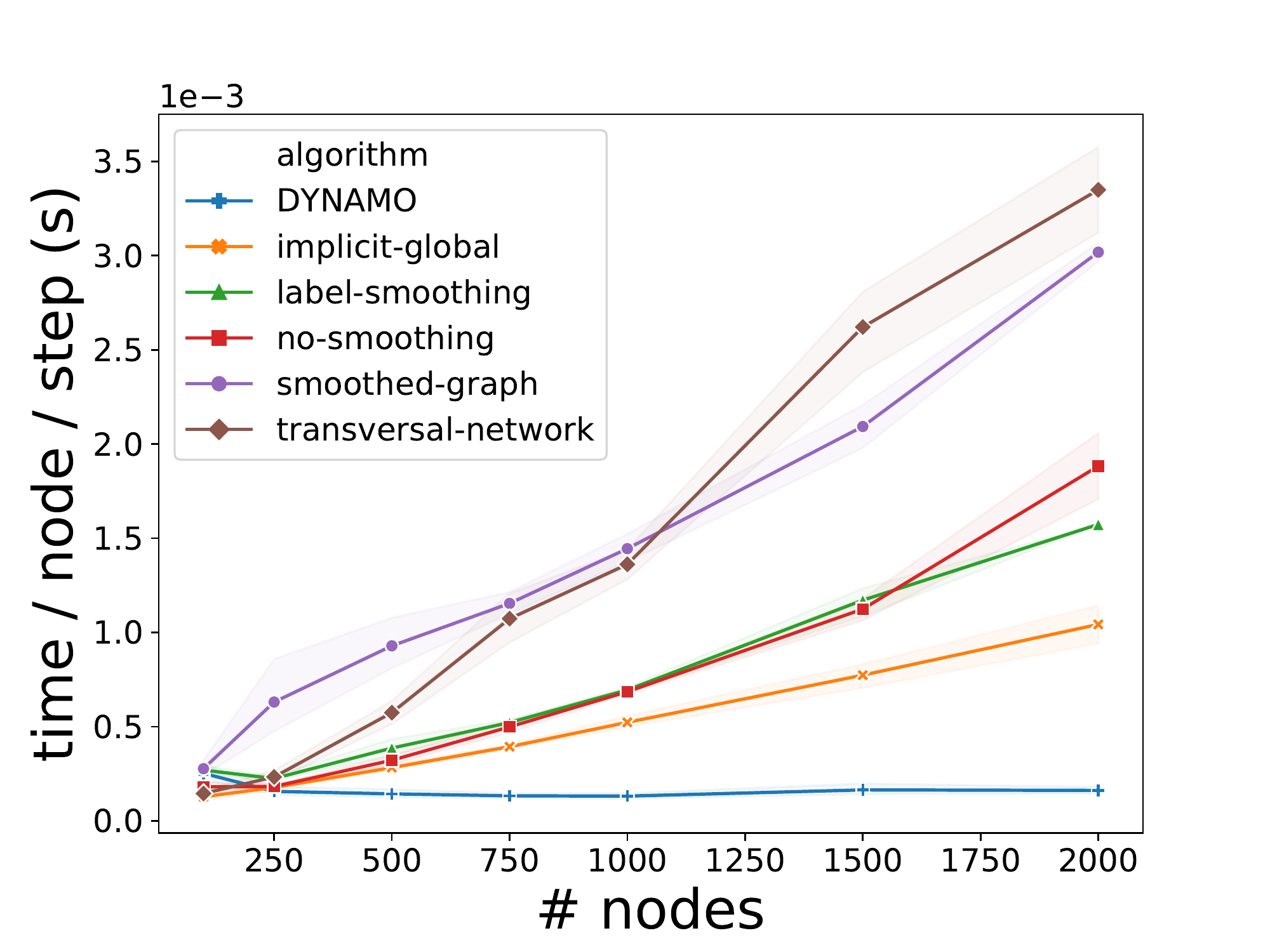}
    \caption{Scalability for different size of graphs, for a same duration (50 steps)}
  \end{subfigure}

    \caption{Running time (in seconds) of the different algorithms depending on the average size of graphs and their number of steps. DYNAMO is by far the fastest method, thanks to its incremental approach. We observe that label smoothing is the only approach whose running time at each step grows with the number of steps.}
  \label{fig:scalability}
\end{figure}

In Fig.\ref{fig:scalability}, we can observe that DYNAMO is the fastest method when the number of nodes or steps becomes large. This result is expected since the method is the only one among those tested that make only \textit{local} changes according to modifications between steps\cite{cazabet2019challenges}.

Most of the tested methods have a complexity that is linear with the number of steps when the graph size is fixed. Label-Smoothing is the exception with complexity increasing quickly with the number of steps, because the similarity of communities in every step needs to be computed with communities in all other steps.

When the size of networks increase, all methods %expect
but DYNAMO have a complexity increasing faster than the number of nodes, which is expected since even the Louvain algorithm has a complexity superior to $O(n)$.

In particular, Transversal-Network and Smoothed-Graph approaches quickly become prohibitive. A possible explanation is that both methods work with dense matrices, due to the computation of an intermediate network representation in which a dense matrix is subtracted to the adjacency matrix. Optimized implementations could partly solve those problems.

Obviously, those results are highly dependent on the implementation of each method, and on the number of available cores for parallelization. For DYNAMO and transversal-network, we used implementations provided by the authors, respectively in Java for the first, and Matlab for the other. For the other methods, we used our own implementation in python, relying on the networkx \cite{hagberg2008exploring}, CDlib \cite{rossetti2019cdlib} and sklearn \cite{scikit-learn} libraries. We run the code on a 4 cores, 16GB of RAM computer.

From those observations, we can conclude that all tested algorithms but DYNAMO are able to handle small graphs with a few thousand steps, or larger graphs with a few hundred steps, but are not designed to handle large graphs with many steps of evolution, due to the handling of dynamic graphs as a succession of snapshots.

%\section{Discussion}
% \paragraph{Advantages and drawbacks of the DSABM}
%Since edges at any step does not depend on previous partitions, the Random Internal Structure property is preserved. And because pairs of nodes with a high Latent Affinity are more likely to be chosen, it ensure a reasonable compliance with the Economy of change property. As a downside, it introduces effects that could affect negatively some algorithms by always preserving edges of highest Latent Affinities, independently of the affiliation of their endpoints. We consider this bias an acceptable drawback, and note that the assumption of an underlying Lattent Affinity could be justified for several kinds of networks, such as social ones.
%
%Another drawback is the complete stability of edges, which might be realistic to model dynamic graph of \textit{relations} (e.g., followers/friends in an online social network) but not for graphs of \textit{interactions} (e.g. phone calls aggregated at a low temporal granularity). A straightforward modification would be to introduce at each step random edge modifications of the desired intensity.

\section{Conclusion}

We have proposed a benchmark to generate dynamic graphs following any community evolution scenario using an appropriate language, and used it to compare several algorithms. By using examples and quantitative analysis, we have shown the weaknesses and strengths of several approaches, in particular the effect of smoothing on the quality of dynamic communities.

A limitation of the current implementation of the benchmark is the space complexity induced by the latent affinity $\Omega$, which is in $n^2$, thus not practical for large graphs.

Variants of the method could be introduced to test different types of dynamic networks: one could test the influence of taking snapshots at coarser temporal granularity using sliding windows, or generate link streams by associating a spawning probability to edges of the currently generated dynamic graphs. Currently, generated networks are non-oriented and unweighted, but the benchmark could be trivially extended to generate such graphs.

Although we proposed scores specifically designed to evaluate dynamic partitions, the question of which score to use remains an important research question. In this work, we did not take into account the events themselves (\textit{split}, \textit{merge}, etc.) in evaluation scores, and based our longitudinal scores on labels only. We think that this approach is not fully satisfying and a proper way to take into account both the stability of communities and their similarity with a ground truth defined at each step should be investigated further.

%This score can be interpreted as follows: every time a community changes, it loses part of its identity. The more change occurs, the more identity is lost, as encoded in the \textit{identity preservation property}. But this loss is not cumulative, as exemplified by the \textit{ship of Theseus} example. Intuitively, this comes from the fact that

%\begin{figure}
%\center
%  \includegraphics[width=0.8\linewidth]{GT.pdf}
%  \caption{A simple scenario of community evolution. Each horizontal line corresponds to a node. Colors correspond to labels of communities. Grey nodes correspond to no affiliation, and correspond to periods during which events are \textit{on-going}, affected communities not being properly defined. White color corresponds to a node not being present in the dynamic network}
%  \label{fig:GT}
%\end{figure}

\section*{Funding}
This work is partially supported by BITUNAM Project ANR-18-CE23-0004 and by the European Community's H2020 Program under the funding scheme 'INFRAIA-01-2018-2019: Research and Innovation action', Grant Agreement n. 871042 'SoBigData++: European Integrated Infrastructure for Social Mining and Big Data Analytics'.

\bibliographystyle{comnet}

\begin{thebibliography}{00}


  
\bibitem{aynaud2010static}
Aynaud, T. {\&} Guillaume, J.-L. (2010)  Static community detection algorithms
  for evolving networks. In {\em 8th International symposium on modeling and
  optimization in mobile, Ad Hoc, and wireless networks}, pages 513--519. IEEE.

\bibitem{bazzi2016generative}
Bazzi, M., Jeub, L.~G., Arenas, A., Howison, S.~D. {\&} Porter, M.~A. (2016)
  Generative benchmark models for mesoscale structure in multilayer networks.
  {\em arXiv preprint arXiv:1608.06196}.

\bibitem{benyahia2016dancer}
Benyahia, O., Largeron, C., Jeudy, B. {\&} Za{\"\i}ane, O.~R. (2016)  Dancer:
  Dynamic attributed network with community structure generator. In {\em Joint
  European Conference on Machine Learning and Knowledge Discovery in
  Databases}, pages 41--44. Springer.

\bibitem{blondel2008fast}
Blondel, V.~D., Guillaume, J.-L., Lambiotte, R. {\&} Lefebvre, E. (2008)  Fast
  unfolding of communities in large networks. {\em Journal of statistical
  mechanics: theory and experiment}, P10008.

\bibitem{cazabet2019challenges}
Cazabet, R. {\&} Rossetti, G. (2019)  Challenges in community discovery on
  temporal networks. In {\em Temporal Network Theory}, pages 181--197.
  Springer.

\bibitem{chykhradze2014distributed}
Chykhradze, K., Korshunov, A., Buzun, N., Pastukhov, R., Kuzyurin, N.,
  Turdakov, D. {\&} Kim, H. (2014)  Distributed generation of billion-node
  social graphs with overlapping community structure. In {\em Complex Networks
  V}, pages 199--208. Springer.

\bibitem{coppens2019comparative}
Coppens, L., De~Venter, J., Mitrovi, S. {\&} De~Weerdt, J. (2019)  A
  comparative study of community detection techniques for large evolving
  graphs. In {\em LEG@ ECML: The third International Workshop on Advances in
  Managing and Mining Large Evolving Graphs collocated with ECML-PKDD}.
  Springer.

\bibitem{falkowski2007data}
Falkowski, T. {\&} Spiliopoulou, M. (2007)  Data Mining for Community
  Dynamics. {\em KI}, \textbf{21}(3), 23--29.

\bibitem{folino2013evolutionary}
Folino, F. {\&} Pizzuti, C. (2013)  An evolutionary multiobjective approach for
  community discovery in dynamic networks. {\em IEEE Transactions on Knowledge
  and Data Engineering}, \textbf{26}(8), 1838--1852.

\bibitem{Genois2018}
Genois, M. {\&} Barrat, A. (2018)  Can co-location be used as a proxy for
  face-to-face contacts?. {\em EPJ Data Science}, \textbf{7}(1), 11.

\bibitem{ghasemian2016detectability}
Ghasemian, A., Zhang, P., Clauset, A., Moore, C. {\&} Peel, L. (2016)
  Detectability thresholds and optimal algorithms for community structure in
  dynamic networks. {\em Physical Review X}, \textbf{6}(3), 031005.

\bibitem{granell2015benchmark}
Granell, C., Darst, R.~K., Arenas, A., Fortunato, S. {\&} G{\'o}mez, S. (2015)
  Benchmark model to assess community structure in evolving networks. {\em
  Physical Review E}, \textbf{92}(1), 012805.

\bibitem{greene2010tracking}
Greene, D., Doyle, D. {\&} Cunningham, P. (2010)  Tracking the evolution of
  communities in dynamic social networks. In {\em Advances in social networks
  analysis and mining (ASONAM), 2010 international conference on}, pages
  176--183. IEEE.

\bibitem{guo2014evolutionary}
Guo, C., Wang, J. {\&} Zhang, Z. (2014)  Evolutionary community structure
  discovery in dynamic weighted networks. {\em Physica A: Statistical Mechanics
  and its Applications}, \textbf{413}, 565--576.

\bibitem{hagberg2008exploring}
Hagberg, A., Swart, P. {\&} S~Chult, D. (2008)  Exploring network structure,
  dynamics, and function using NetworkX. Technical report, Los Alamos National
  Lab.(LANL), Los Alamos, NM (United States).

\bibitem{holland1983stochastic}
Holland, P.~W., Laskey, K.~B. {\&} Leinhardt, S. (1983)  Stochastic
  blockmodels: First steps. {\em Social networks}, \textbf{5}(2), 109--137.

\bibitem{kawadia2012sequential}
Kawadia, V. {\&} Sreenivasan, S. (2012)  Sequential detection of temporal
  communities by estrangement confinement. {\em Scientific reports},
  \textbf{2}, 794.

\bibitem{kobayashi2019structured}
Kobayashi, T., Takaguchi, T. {\&} Barrat, A. (2019)  The structured backbone of
  temporal social ties. {\em Nature communications}, \textbf{10}(1), 1--11.

\bibitem{lancichinetti2008benchmark}
Lancichinetti, A., Fortunato, S. {\&} Radicchi, F. (2008)  Benchmark graphs for
  testing community detection algorithms. {\em Physical review E},
  \textbf{78}(4), 046110.

\bibitem{leskovec2005graphs}
Leskovec, J., Kleinberg, J. {\&} Faloutsos, C. (2005)  Graphs over time:
  densification laws, shrinking diameters and possible explanations. In {\em
  Proceedings of the eleventh ACM SIGKDD international conference on Knowledge
  discovery in data mining}, pages 177--187.

\bibitem{snapnets}
Leskovec, J. {\&} Krevl, A. (2014)  {SNAP Datasets}: {Stanford} Large Network
  Dataset collection. \url{http://snap.stanford.edu/data}.

\bibitem{li2020optimization}
Li, H. J., Wang, L., Zhang, Y., {\&} Perc  M. (2020). Optimization of identifiability for efficient community detection. In {\em New Journal of Physics, 22(6)}.

\bibitem{lin2008facetnet}
Lin, Y.-R., Chi, Y., Zhu, S., Sundaram, H. {\&} Tseng, B.~L. (2008)  Facetnet:
  a framework for analyzing communities and their evolutions in dynamic
  networks. In {\em Proceedings of the 17th international conference on World
  Wide Web}, pages 685--694. ACM.

\bibitem{linhares2017dynetvis}
Linhares, C.~D., Traven{\c{c}}olo, B.~A., Paiva, J. G.~S. {\&} Rocha, L.~E.
  (2017)  DyNetVis: A system for visualization of dynamic networks. In {\em
  Proceedings of the Symposium on Applied Computing}, pages 187--194. ACM.

\bibitem{mucha2010community}
Mucha, P.~J., Richardson, T., Macon, K., Porter, M.~A. {\&} Onnela, J.-P.
  (2010)  Community structure in time-dependent, multiscale, and multiplex
  networks. {\em science}, \textbf{328}(5980), 876--878.

\bibitem{scikit-learn}
Pedregosa, F., Varoquaux, G., Gramfort, A., Michel, V., Thirion, B., Grisel,
  O., Blondel, M., Prettenhofer, P., Weiss, R., Dubourg, V., Vanderplas, J.,
  Passos, A., Cournapeau, D., Brucher, M., Perrot, M. {\&} Duchesnay, E. (2011)
   Scikit-learn: Machine Learning in {P}ython. {\em Journal of Machine Learning
  Research}, \textbf{12}, 2825--2830.

\bibitem{perc2010coevolutionary}
Perc, M. {\&} Szolnoki, A. (2010). Coevolutionary games—a mini review. In {\em BioSystems, 99(2)} pages109-125.

\bibitem{Rossetti2017}
Rossetti, G. (2017)  $\text{RD}\small{\text{YN}}$: graph benchmark handling
  community dynamics. {\em Journal of Complex Networks}, \textbf{5}(6),
  893--912.

\bibitem{rossetti2018community}
Rossetti, G. {\&} Cazabet, R. (2018)  Community discovery in dynamic networks:
  a survey. {\em ACM Computing Surveys (CSUR)}, \textbf{51}(2), 1--37.

\bibitem{rossetti2019cdlib}
Rossetti, G., Milli, L. {\&} Cazabet, R. (2019)  CDLIB: a python library to
  extract, compare and evaluate communities from complex networks. {\em Applied
  Network Science}, \textbf{4}(1), 52.

\bibitem{sarzynska2015null}
Sarzynska, M., Leicht, E.~A., Chowell, G. {\&} Porter, M.~A. (2015)  Null
  models for community detection in spatially embedded, temporal networks. {\em
  Journal of Complex Networks}, \textbf{4}(3), 363--406.

\bibitem{sengupta2017benchmark}
Sengupta, N., Hamann, M. {\&} Wagner, D. (2017)  Benchmark Generator for
  Dynamic Overlapping Communities in Networks. In {\em Data Mining (ICDM), 2017
  IEEE International Conference on}, pages 415--424. IEEE.

\bibitem{tantipathananandh2011finding}
Tantipathananandh, C. {\&} Berger-Wolf, T.~Y. (2011)  Finding communities in
  dynamic social networks. In {\em Data Mining (ICDM), 2011 IEEE 11th
  International Conference on}, pages 1236--1241. IEEE.

\bibitem{xu2014dynamic}
Xu, K.~S. {\&} Hero, A.~O. (2014)  Dynamic stochastic blockmodels for
  time-evolving social networks. {\em IEEE Journal of Selected Topics in Signal
  Processing}, \textbf{8}(4), 552--562.

\bibitem{zhang2017random}
Zhang, X., Moore, C. {\&} Newman, M.~E. (2017)  Random graph models for dynamic
  networks. {\em The European Physical Journal B}.

\bibitem{zhuang2019dynamo}
Zhuang, D., Chang, M. J., {\&} Li, M. (2019). DynaMo: Dynamic community detection by incrementally maximizing modularity. {\em IEEE Transactions on Knowledge and Data Engineering.}

\end{thebibliography}

\newpage

\appendix
\section{Appendix: Definition of community events}
\label{events_definition}

In this appendix, we give the implementation of several well-known community events. Note that these events, and a few more, are implemented in the library released with this article.

\textbf{BIRTH(NB-NODES, LABEL)} is an instruction to create a new community with a label LABEL composed of a number of NB-NODES newly created nodes. Note that a new community could appear by taking nodes from one or several existing communities, but this should be represented through an ad-hoc ASSIGN event.

\begin{algorithm2e}[H]
\SetKw{KwIs}{is}
\SetKw{where}{where}

\KwIn{NB-NODES, LABEL}
\Begin{
N $\gets$ []\;
\For{i $\gets$ 0  $\KwTo$ NB-NODES}{
	N $\gets$ N $\cup$ NEW-NODE() \;
}
 \Return ASSIGN([],[N],[LABEL]) \;
}

\caption{instruction \textbf{BIRTH}}
\end{algorithm2e}
\vspace{1cm}

\textbf{DEATH(COM)} is an instruction to make the community COM disappear, and its nodes leave the network.

\begin{algorithm2e}[H]
\SetKw{KwIs}{is}
\SetKw{where}{where}

\KwIn{$COM$}
\Begin{
%$C \gets <\_,b,n> \in AC \where b=l$ \;
	ASSIGN([COM],[],[]) \;
}
\caption{instruction \textbf{DEATH}}
  \end{algorithm2e}
\vspace{1cm}

\textbf{MERGE(C-BEFORE, L-AFTER)} is used to merge two or more communities into a single one. It is an instruction with two parameters:
\begin{itemize}
	\item C-BEFORE the list of communities to merge
	\item L-AFTER the label of the resulting community (if not provided, a new unique label is generated)
\end{itemize}

The label in L-AFTER can be either one of those of the communities in C-BEFORE or a new one (conservation or not of the label by one of the resulting communities).
It is defined by the following algorithm:

\begin{algorithm2e}[H]
\KwIn{C-BEFORE, L-AFTER}
\SetKw{BIRTH}{BIRTH}
\SetKw{MIGRATE}{MIGRATE}
\SetKw{DEATH}{DEATH}

\Begin{
N $\gets$ []\;
\For{C $\in$ C-BEFORE}{
	N $\gets$ N $\cup$ C.N \;
}

\Return ASSIGN(C-BEFORE,[N],[L-AFTER]) \;

}
\caption{instruction \textbf{MERGE}}
  \end{algorithm2e}
\vspace{0.5cm}

\textbf{SPLIT(C-BEFORE, L-AFTER, SIZES)} is used to split a community by creating new ones of sizes described in the SIZE parameter. Nodes are chosen randomly. It can be defined as a command with three parameters:
\begin{itemize}
	\item C-BEFORE: the community to split
	\item L-AFTER: the ordered list of labels of the communities created by the split event
	\item SIZES: an ordered list containing the sizes of the resulting communities. The sum of the elements
	of this
	list must be equal to the number of nodes in C-BEFORE

\end{itemize}

It is defined by the following algorithm:

\begin{algorithm2e}[H]
\SetKw{KwTo}{to}
\SetKw{length}{length}
\SetKw{randomChoice}{randomChoice}

\KwIn{C-BEFORE, SIZES, L-AFTER}

\Begin{
N $\gets$  C-BEFORE.N \;
AFTER-NODES $\gets$ [] \;
\For{i $\gets$ 0 $\KwTo$ $\length$(SIZES)}{
 	AFTER-NODES[i] $\gets$ $\randomChoice$(N, SIZES[i]) \;
 	N $\gets$ N $\setminus$  AFTER-NODES[i] \;
}

\Return ASSIGN([C-BEFORE], AFTER-NODES, L-AFTER) \;

}
\caption{instruction \textbf{SPLIT}. randomChoice(L,n) function select randomly n elements in list L }
  \end{algorithm2e}

\vspace{0.5cm}

\textbf{THESEUS(COM, NB-NODES)}
COM is the community to modify. NB-NODES corresponds to the number of nodes to replace in the original community. If NB-NODES is equal to the number of nodes, then the event corresponds exactly to the scenario described in fig. \ref{fig:ship}.

Note that we use the ASSIGN instruction to describe, at each step, that the community simultaneously lose a node and gain a new one, and then at the end to make the community reappear with the original nodes.

\noindent
%\underline{Algorithm}

\begin{algorithm2e}[H]
\SetKw{KwTo}{to}
\SetKw{KwIs}{is}
\SetKw{randomChoice}{randomChoice}

\SetKw{where}{where}
\KwIn{COM, NB-NODES}

\Begin{
  N-CURRENT $\gets$ copy(COM.N)\;
  N-ORIGINAL $\gets$ copy(COM.N) \;
  COM-CURRENT $\gets$ COM \;

  \For{i $\gets$ 0 $\KwTo$ NB-NODES}{
  REMOVED $\gets$ $\randomChoice$(N-CURRENT,1) \;
  N-CURRENT $\gets$ N-CURRENT $\setminus$ REMOVED \;
  ADDED $\gets$ NEW-NODE() \;
  [COM-CURRENT] $\gets$ ASSIGN([COM-CURRENT], [(COM-CURRENT.N $\cup$ ADDED) $\setminus$ REMOVED ], [COM-CURRENT.L]) \;
  %[COM-CURRENT] $\gets$ ASSIGN([COM-CURRENT], [N-CURRENT.N $\cup$ ADDED], [COM-CURRENT.L]) \;
  }

  B $\gets$ ASSIGN([], [N-ORIGINAL], [NEW-COM-ID()]) \;
  \Return $[$COM-CURRENT, B$]$
}
\caption{instruction \textbf{THESEUS}. copy function allows to make a copy of a list, such as it can be modified without affecting the original}
  \end{algorithm2e}

\vspace{0.5cm}

\textbf{INITIALIZE(SIZES, LABELS)} allows setting the community structure in the first step.
\begin{itemize}
	\item SIZES is the list of sizes of initial communities
	\item LABELS is the list of labels of those communities
\end{itemize}
  The function returns the corresponding communities composed of newly created nodes.
The code is omitted for simplicity, but present in the provided implementation.

\end{document}